\documentclass[twoside,twocolumn,9pt]{article}
\usepackage{extsizes}
\usepackage[super,sort&compress,comma]{natbib} 
\usepackage[version=3]{mhchem}
\usepackage[left=1.5cm, right=1.5cm, top=1.785cm, bottom=2.0cm]{geometry}
\usepackage{balance}
\usepackage{mathptmx}
\usepackage{sectsty}
\usepackage{graphicx} 
\usepackage{lastpage}
\usepackage[format=plain,justification=justified,singlelinecheck=false,font={stretch=1.125,small,sf},labelfont=bf,labelsep=space]{caption}
\usepackage{float}
\usepackage{fancyhdr}
\usepackage{fnpos}
\usepackage[english]{babel}
\addto{\captionsenglish}{%
  \renewcommand{\refname}{Notes and references}
}
\usepackage{array}
\usepackage{droidsans}
\usepackage{charter}
\usepackage[T1]{fontenc}
\usepackage[usenames,dvipsnames]{xcolor}
\usepackage{setspace}
\usepackage[compact]{titlesec}
\usepackage{hyperref}
\usepackage{subcaption}
\usepackage{siunitx}
\usepackage{soul}
\usepackage{enumitem}
\definecolor{cream}{RGB}{222,217,201}

\begin{document}

\pagestyle{fancy}
\thispagestyle{plain}
\fancypagestyle{plain}{
%%%HEADER%%%
\renewcommand{\headrulewidth}{0pt}
}
%%%END OF HEADER%%%

%%%PAGE SETUP - Please do not change any commands within this section%%%
\makeFNbottom
\makeatletter
\renewcommand\LARGE{\@setfontsize\LARGE{15pt}{17}}
\renewcommand\Large{\@setfontsize\Large{12pt}{14}}
\renewcommand\large{\@setfontsize\large{10pt}{12}}
\renewcommand\footnotesize{\@setfontsize\footnotesize{7pt}{10}}
\makeatother

\renewcommand{\thefootnote}{\fnsymbol{footnote}}
\renewcommand\footnoterule{\vspace*{1pt}% 
\color{cream}\hrule width 3.5in height 0.4pt \color{black}\vspace*{5pt}} 
\setcounter{secnumdepth}{5}

\makeatletter 
\renewcommand\@biblabel[1]{#1}            
\renewcommand\@makefntext[1]% 
{\noindent\makebox[0pt][r]{\@thefnmark\,}#1}
\makeatother 
\renewcommand{\figurename}{\small{Fig.}~}
\sectionfont{\sffamily\Large}
\subsectionfont{\normalsize}
\subsubsectionfont{\bf}
\setstretch{1.125} %In particular, please do not alter this line.
\setlength{\skip\footins}{0.8cm}
\setlength{\footnotesep}{0.25cm}
\setlength{\jot}{10pt}
\titlespacing*{\section}{0pt}{4pt}{4pt}
\titlespacing*{\subsection}{0pt}{15pt}{1pt}
%%%END OF PAGE SETUP%%%

%%%FOOTER%%%
\fancyfoot{}
\fancyfoot[LO,RE]{\vspace{-7.1pt}\includegraphics[height=9pt]{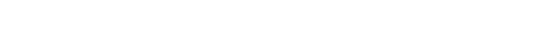}}
%\fancyfoot[CO]{\vspace{-7.1pt}\hspace{13.2cm}\includegraphics{head_foot/RF}}
%\fancyfoot[CE]{\vspace{-7.2pt}\hspace{-14.2cm}\includegraphics{head_foot/RF}}
\fancyfoot[RO]{\footnotesize{\sffamily{1--\pageref{LastPage} ~\textbar  \hspace{2pt}\thepage}}}
\fancyfoot[LE]{\footnotesize{\sffamily{\thepage~\textbar\hspace{3.45cm} 1--\pageref{LastPage}}}}
\fancyhead{}
\renewcommand{\headrulewidth}{0pt} 
\renewcommand{\footrulewidth}{0pt}
\setlength{\arrayrulewidth}{1pt}
\setlength{\columnsep}{6.5mm}
\setlength\bibsep{1pt}
%%%END OF FOOTER%%%

%%%FIGURE SETUP - please do not change any commands within this section%%%
\makeatletter 
\newlength{\figrulesep} 
\setlength{\figrulesep}{0.5\textfloatsep} 

\newcommand{\topfigrule}{\vspace*{-1pt}% 
\noindent{\color{cream}\rule[-\figrulesep]{\columnwidth}{1.5pt}} }

\newcommand{\botfigrule}{\vspace*{-2pt}% 
\noindent{\color{cream}\rule[\figrulesep]{\columnwidth}{1.5pt}} }

\newcommand{\dblfigrule}{\vspace*{-1pt}% 
\noindent{\color{cream}\rule[-\figrulesep]{\textwidth}{1.5pt}} }

\makeatother
%%%END OF FIGURE SETUP%%%

%%%TITLE, AUTHORS AND ABSTRACT%%%
\twocolumn[
  \begin{@twocolumnfalse}
%{\includegraphics[height=30pt]{head_foot/journal_name}\hfill\raisebox{0pt}[0pt][0pt]{\includegraphics[height=55pt]{head_foot/RSC_LOGO_CMYK}}\\[1ex]
%\includegraphics[width=18.5cm]{head_foot/header_bar}}\par
\vspace{1em}
\sffamily
%\begin{tabular}{m{4.5cm} p{13.5cm} }
\begin{tabular}{m{18cm}}

%\includegraphics{head_foot/DOI} & \noindent\LARGE{\textbf{Performance  of the
%nanopost single-photon source: beyond the single-mode model}} \\%Article title goes here instead of the text "This is the title"
%\LARGE{\textbf{Breakdown of the single-mode Fabry-Pérot model for the nanopost single-photon source$^\dag$}} \\%Article title goes here instead of the text "This is the title"
\LARGE{\textbf{Performance  of the
nanopost single-photon source: beyond the single-mode model}}
\vspace{0.3cm} \\

 \noindent\large{Martin Arentoft Jacobsen,$^{\ast}$\textit{$^{a}$} Yujing Wang,\textit{$^{a}$} Luca Vannucci,\textit{$^{a}$} Julien Claudon\textit{$^{b}$}, Jean-Michel Gérard,\textit{$^{b}$} and Niels Gregersen\textit{$^{a}$}} \\%Author names go here instead of "Full name", etc.
\vspace{0.3cm}
\normalsize{We present a detailed analysis of the physics governing the collection efficiency and the Purcell enhancement of the nanopost single-photon source. We show that a standard single-mode Fabry-Pérot model is insufficient to describe the device performance, which benefits significantly from scattering from the fundamental mode to radiation modes. We show how the scattering mechanism decouples the collection efficiency from the Purcell enhancement, such that maximum collection efficiency is obtained off-resonance. Finally, we discuss how this scattering mechanism can be beneficial for future single-photon source designs.} \\

%\normalsize{We present a detailed physical analysis of the breakdown of the single-mode Fabry-Pérot model for the nanopost single-photon source. We show that the breakdown is caused by strong scattering into radiation of the fundamental mode, which significantly increases the collection efficiency, and decouples the collection efficiency and the Purcell factor. Finally, we discuss how this scattering mechanism can be beneficial for future single-photon source designs.} \\

%The abstrast goes here instead of the text "The abstract should be..."
%The abstract should be a single paragraph which summarises the content of the article. Any references in the abstract should be written out in full \textit{e.g.}\ [Surname \textit{et al., Journal Title}, 2000, \textbf{35}, 3523].
\end{tabular}

 \end{@twocolumnfalse} \vspace{0.6cm}

  ]
%%%END OF TITLE, AUTHORS AND ABSTRACT%%%

%%%FONT SETUP - please do not change any commands within this section
\renewcommand*\rmdefault{bch}\normalfont\upshape
\rmfamily
\section*{}
\vspace{-1cm}

%%%FOOTNOTES%%%

\footnotetext{\textit{$^{a}$~DTU Electro, Department of Electrical and Photonics Engineering, Technical University of Denmark, DK-2800 Kongens Lyngby, Denmark. E-mail: maaja@dtu.dk}}
\footnotetext{\textit{$^{b}$~Univ. Grenoble Alpes, CEA, Grenoble INP, IRIG, PHELIQS, “Nanophysique et Semiconducteurs” Group, F-38000 Grenoble, France. }}

%Please use \dag to cite the ESI in the main text of the article.
%If you article does not have ESI please remove the the \dag symbol from the title and the footnotetext below.

%%%END OF FOOTNOTES%%%

%%%MAIN TEXT%%%%
\section{Introduction}

The construction of scalable optical quantum technologies \cite{Pan2012,OBrien2009} relies on the development of sources of single indistinguishable photons \cite{Arakawa2020, Aharonovich2016, Gregersen2013, Gregersen2017} and of entangled photon pairs \cite{Huber2018}. The ideal single-photon source (SPS) should be deterministic and feature pure emission of single photons. The main figure of merit \cite{Gregersen2017} is the collection efficiency $\varepsilon$ defined as the number of photons detected in the out-coupling channel per trigger. In a multi-photon interference experiment \cite{Wang2019c} with $N$ photons, the success probability $P$ scales as $P=\varepsilon ^N$, and increasing $\varepsilon$ towards 1 is thus critical to achieve scalable optical quantum information processing. The spontaneous parametric down conversion process \cite{Kwiat1995} is a straight-forward technique widely used within the quantum optics community for production of highly indistinguishable photons, however its probabilistic nature limits the efficiency of pure photon emission to a few percent. 

For this reason, the community has turned its attention towards two level systems, in particular the semiconductor quantum dot \cite{Arakawa2020,Aharonovich2016,Shields2009} (QD), capable of deterministic emission of single photons. For a QD in a bulk material, $\varepsilon$ is limited to a few percent this time due to the large index contrast at the semiconductor-air interface. It is thus necessary to place the QD inside a photonic nanostructure \cite{Gregersen2013,Gregersen2017} directing the light towards the collection optics. A main strategy for controlling the light emission is to place the QD inside a micro cavity and exploit cavity quantum electrodynamics (CQED) in the weak coupling regime to selectively enhance the light emission into the optical mode of the microcavity using the Purcell effect \cite{JMG1998}. 
Detailed understanding of the CQED physics governing the collection efficiency can be obtained using a single-mode Fabry-Pérot description \cite{Friedler2009, Gregersen2016,  Wang2020_PRB_Biying} of the light emission. Here, the spontaneous emission $\beta$ factor describes the emission rate $\Gamma_{\rm C}$ of the QD into a fundamental HE$_{11}$ cavity mode divided by the total emission rate $\Gamma_{\rm T} = \Gamma_{\rm C} + \Gamma_{\rm B}$ including a contribution $\Gamma_{\rm B}$ to background radiation modes. The rate $\Gamma_{\rm C}$ into the cavity mode normalized to the rate $\Gamma_{\rm Bulk}$ in a bulk medium is quantified by the Purcell \cite{Purcell1946} factor $F_{\rm p} = \Gamma_{\rm C}/\Gamma_{\rm Bulk} = \frac{3}{4 \pi^2} \frac{Q}{V_{\rm n}}$ at resonance, where $Q$ is the cavity quality factor and  $V_{\rm n}$ is the mode volume in units of material cubic wavelengths $(\lambda/n)^3$. The spontaneous emission $\beta$ factor can then be written in terms of the Purcell factor as 
\begin{align}
	\beta = \frac{\Gamma_{\rm C}}{\Gamma_{\rm C} +  \Gamma_{\rm B}} = \frac{F_{\rm p} }{F_{\rm p}  + \Gamma_{\rm B} / \Gamma_{\rm Bulk}}.
	\label{beta_eq}
\end{align}
Furthermore, we define the transmission $\gamma$ as the fraction of power in the cavity mode detected by the collection optics. Finally, we can then define a single-mode Fabry-Pérot model (SMM) $\varepsilon _{\rm s}$ for the efficiency as $\varepsilon _{\rm s} = \beta \gamma $. From Eq.\ \eqref{beta_eq}, we observe that increasing the Purcell factor $F_{\rm p}$ will improve the collection efficiency, and maximum efficiency is thus expected for a QD on resonance with the cavity. 

Indeed, this design paradigm that Purcell enhancement is beneficial for achieving high collection efficiency is well-established within SPS engineering: The most succesful SPS design strategies today include the microcavity pillar \cite{Wang2019b, Wang2020_PRB_Biying, Somaschi2016, Ding2016} and the open cavity approach \cite{Tomm2021} demonstrating up to $\varepsilon \sim$~0.6 into a first lens \cite{Wang2019b} and into a fiber \cite{Tomm2021}, respectively, combined with highly indistinguishable photon emission. These narrowband approaches, for which the single mode model $\varepsilon _{\rm s} = \beta \gamma $ is an excellent approximation \cite{Wang2020_PRB_Biying}, rely critically on resonant Purcell enhancement and thus on control of the spectral alignment \cite{Somaschi2016} to achieve high efficiency.
On the other hand, broadband approaches including the photonic nanowire \cite{Bleuse2011, Gregersen2016, Gaal2022, Claudon2010, Claudon2013} and the photonic crystal waveguide \cite{Lecamp2007b, MangaRao2007, Arcari2014} designs exploit suppression of the background emission rate using the dielectric screening effect \cite{Bleuse2011, Lecamp2007b, MangaRao2007} to non-resonantly maximize the $\beta$ factor. Even so, these broadband approaches also benefit from resonant cavity \cite{Gregersen2016, Gaal2022} and slow-light \cite{Lecamp2007b, MangaRao2007} effects to further improve the efficiency, confirming again that Purcell enhancement is beneficial in the SPS engineering.

\begin{figure}[h]
	%\advance\leftskip-4cm
	\begin{subfigure}{1\linewidth}
		\centering
		\includegraphics[width= 1 \linewidth]{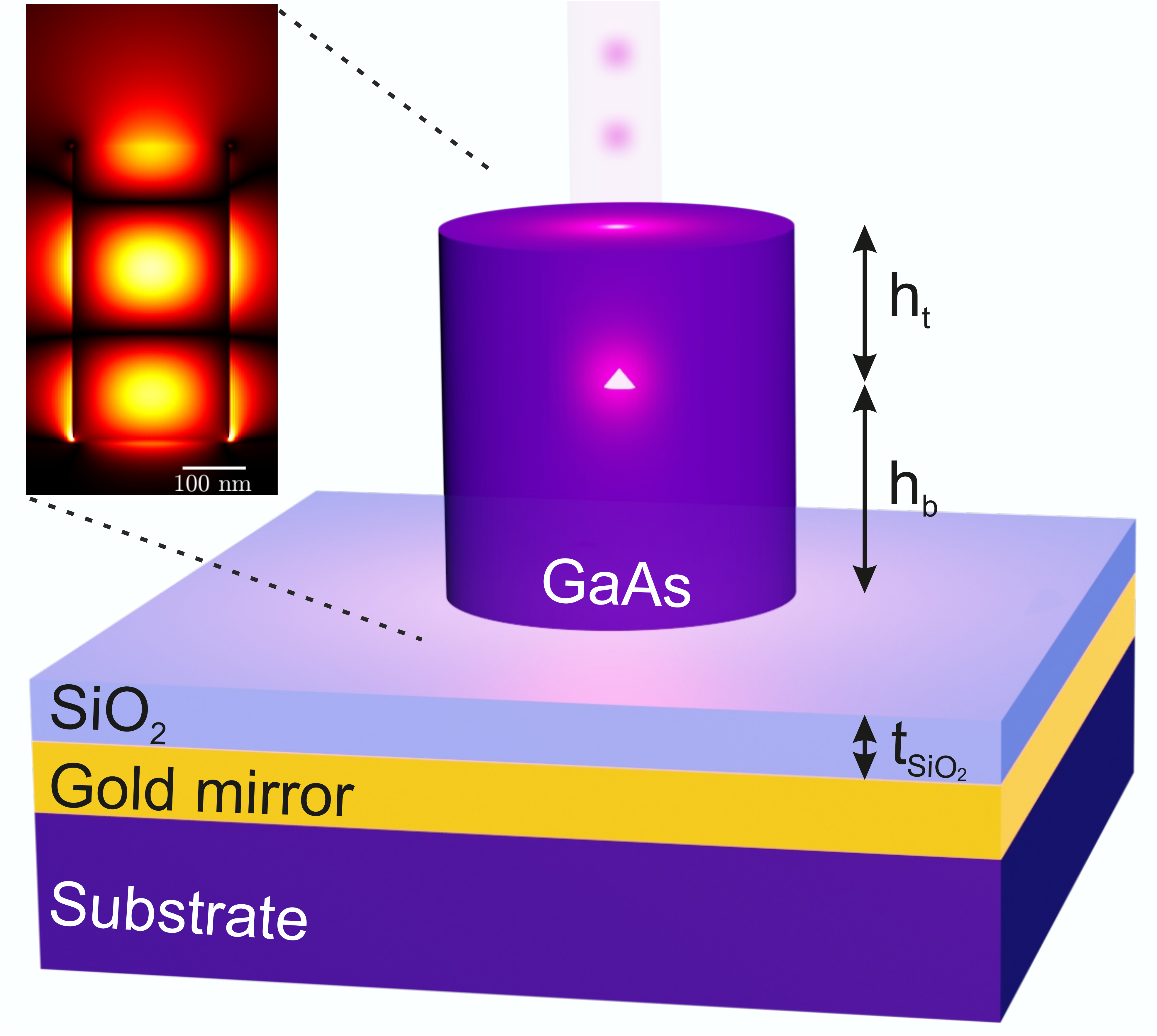}
		%\includegraphics[width= 1 \linewidth]{test_sketch2.pdf}

		%\caption{}
	\end{subfigure}
	\caption{Sketch of the "nanopost" nanowire optical nanocavity and the geometrical parameters. The resulting in-plane electrical field profile of a QD placed inside the nanopost is shown as an inset. The length of the white scale bar in the inset is \SI{100}{nm}.}
	\label{Sketch}
\end{figure}

However, this paradigm has been challenged by new broadband SPS geometries, such as the circular Bragg grating or "bullseye" design \cite{Yao2018, Liu2019, Wang2019b}, for which high collection efficiency is obtained in a wavelength range significantly broader \cite{Yao2018} ($\sim$ 100 nm) than the typical resonance linewidth ($\sim$ 10 nm). Similar characteristics were observed very recently for the nanowire optical nanocavity or "nanopost" design \cite{Kotal2021} shown in Fig.\ (\ref{Sketch}), for which a significant Purcell factor $F_{\rm p}$ of  5.6 enabled by the ultrasmall mode volume of the nanocavity was experimentally demonstrated. Additionally, a surprisingly high collection efficiency of 0.35 was measured \cite{Kotal2021}, which was attributed to a breakdown \cite{AndreasPHD} of the single mode model $\varepsilon _{\rm s} = \beta \gamma$ for the efficiency.

In this work, we investigate this surprising breakdown by performing a detailed quantitative analysis of the physics governing the nanopost geometry. We show that the single mode model fails to describe the physics of both the Purcell enhancement and the collection efficiency due to a decoupling between the two: The computed efficiency is significantly higher than the prediction of the single-mode model thanks to additional transmission channels to the far-field, whose beneficial contributions are dominating over the resonant cavity effect. We show not only that maximum Purcell enhancement and maximum collection efficiency are obtained for entirely different design parameters, but also that maximum efficiency is obtained off-resonance. The analysis is performed using a Fourier Modal Method \cite{Gur2021a}, allowing for direct insight into the beneficial interplay beyond the single-mode model with the continuum of radiation modes.

This article is organized as follows: In Section \ref{Sec:Nanopost}, we present the nanopost and its performance in terms of $F_{\rm p}$ and $\varepsilon$, and we demonstrate the breakdown of the single-mode model. In Section \ref{Sec:Theory}, we present our theoretical framework based on the Fourier Modal Method, which we subsequently use to analyze the complex interplay with radiation mode channels in Section \ref{Sec:Analysis} and its influence on the collection efficiency and the Purcell factor. In Section \ref{Sec:Discussion} we put the nanopost physics into perspective and discuss its impact on SPS engineering, followed by our conclusion. Additional simulation results are presented in the %Appendix / 
Supplementary Information.

\section{The nanopost geometry and the breakdown of the single-mode Fabry-Pérot model}  \label{Sec:Nanopost}

The nanopost shown in Fig.\ (\ref{Sketch}) consists of a truncated GaAs nanowire with diameter $D$ on top of a \ce{SiO2}-\ce{Au} mirror. The top of the nanowire is flat, and the surrounding medium is air. The \ce{SiO2} layer, located between the nanowire and the gold, has a thickness indicated by $t_{\rm SiO_2}$. The QD, modelled as a dipole, is placed on-axis inside the nanowire at a position $h_{\rm b}$ from the bottom interface and $h_{\rm t}$ from the top interface. The refractive indices of the materials are chosen as $n_{\rm GaAs}=3.46$, $n_{\rm SiO_2}=1.5$ and $n_{\rm Au}=0.201+5.85i$ at $\lambda=\SI{930}{nm}$ and assumed to be constant as a function of wavelength.
In the inset of  Fig.\ (\ref{Sketch}), a dipole with an emission wavelength of $\SI{930}{nm}$ is placed inside the nanopost, and the resulting in-plane electrical field, simulated using the FMM, is shown. Three antinodes can be seen in the field profile, corresponding to the order 3 cavity mode, and they are enumerated from the bottom mirror as the 1st, 2nd and 3rd antinode. The field profile generated by the dipole is independent of the vertical position of the dipole, only the intensity changes. The intensity is not the same at the three antinodes due to the breakdown of the SMM.

%- 1 antinode cavity - poor what?
%- 2 antinode cavity - poor far-field?
%- 4+ antinode cavity - approaching 1D model?

We now present the performance of the nanopost as a function of the diameter, $D$, and the silica layer thickness, $t_{\rm SiO_2}$. We have scanned the parameter ranges $D=\SI{196}{nm}$ to $D=\SI{300}{nm}$ and $t_{\rm SiO_2}=\SI{0}{nm}$ to $t_{\rm SiO_2}=\SI{25}{nm}$ and chosen a design wavelength of $\lambda_{\rm d}=\SI{930}{nm}$. The height of the structure and the position of the QD are dynamically changed to keep the order 3 cavity mode resonance at $\lambda_{\rm r}=\SI{930}{nm}$ for the QD at the 2nd antinode. The required procedure is presented in Supplementary \ref{heightpro}.  

\begin{figure}[ht]
	%\advance\leftskip-4cm
	\begin{subfigure}{1\linewidth}
		\centering
		\includegraphics[width= 1 \linewidth]{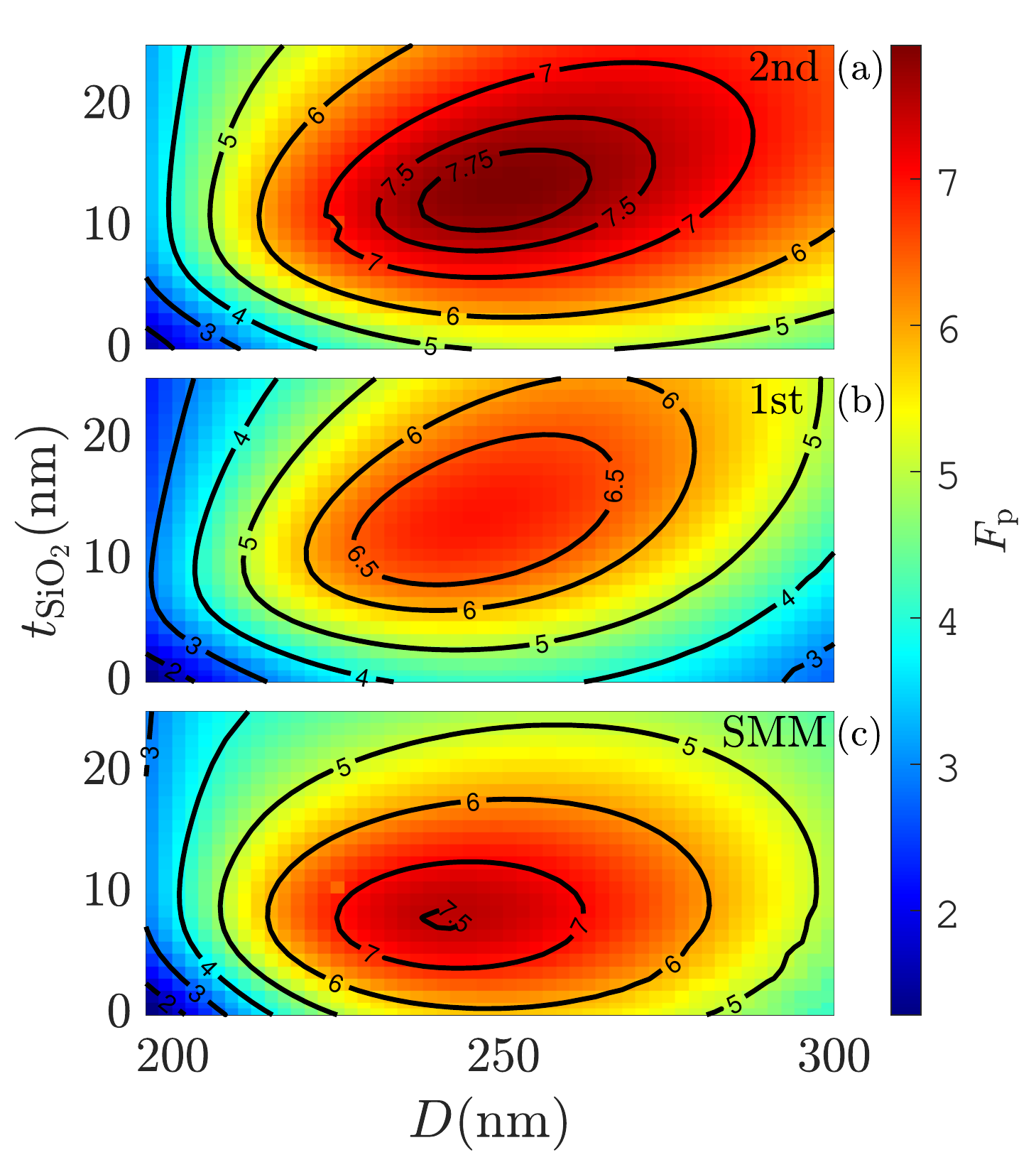}
	\end{subfigure}
%	\caption{(a) $F_{\rm p}$ for the 2nd antinode. (b) $F_{\rm p}$ for the 1st antinode. (c) $F_{\rm p}$ using the 1D model.}
	\caption{Purcell factor, $F_{\rm p}$, for a QD at the 2nd antinode (a), 1st antinode (b) and any antinode using the SMM (c) as a function of the diameter, $D$, and the silica layer thickness, $t_{\rm SiO_2}$.}
	\label{Fp_result}
\end{figure}

The first quantity of interest is the spontaneous emission rate $\Gamma_{\rm T}$. %The normalized total emission rate $\Gamma_{\rm T}/\Gamma_{\rm Bulk}$ differs by $\Gamma_{\rm B}/\Gamma_{\rm Bulk}$ from the Purcell factor $F_{\rm p}$, however 
For the high-$\beta$ structures investigated here, the total normalized rate $\Gamma_{\rm T}/\Gamma_{\rm Bulk}$ and the Purcell factor $F_{\rm p}$ are similar, and we will in the following refer to the normalized total rate as the "Purcell factor $F_{\rm p}$".
In Fig.\ (\ref{Fp_result}a,\ref{Fp_result}b) the Purcell factor is shown as a function of the diameter and the silica layer thickness for a QD placed in the 2nd and 1st antinode. In the entire parameter space, the Purcell factor is larger for the 2nd antinode and a maximum value of $F_{\rm p}=7.9$ is reached at $D=\SI{250}{nm}$ and $t_{\rm SiO_2}=\SI{13}{nm}$. This discrepancy between the two antinodes also demonstrates the deviations of the SMM. Overall, the tendency of the Purcell factor is similar at the 2 antinodes with one peak value. The minimum is located in the corner of no silica and the smallest diameter. In Fig.\ (\ref{Fp_result}c) the Purcell factor is now shown using the SMM. Comparing the SMM to the full model for the two antinodes, there are both positive and negative deviations across most of the parameter space. Compared to the 2nd antinode, the SMM also predicts a slightly lower value for the maximum Purcell factor of $F_{\rm p}=7.5$, but at a very different position of $D=\SI{242}{nm}$ and $t_{\rm SiO_2}=\SI{8}{nm}$. However, compared to the 1st antinode the SMM predicts a larger value for the maxmimum Purcell factor. This is an invitation to obtain a better description and understanding of the physics responsible for the Purcell factor, which we will provide in this paper.

%\begin{figure}[H]
%	\begin{subfigure}{0.5\textwidth}
%		\centering
%		%\includegraphics[width= 1 \textwidth]{GaAsnanoDBR2.eps}
%		\includegraphics[width= 1 \textwidth]{eff2nd.eps}
%		\caption{}
%		\label{eff2nd}
%	\end{subfigure}
%	\begin{subfigure}{0.5\textwidth}
%		\centering
%		%\includegraphics[width= 1 \textwidth]{GaAsnanoDBR2.eps}
%		\includegraphics[width= 1 \textwidth]{eff1st.eps}
%		\caption{}
%		\label{eff1st}
%	\end{subfigure}
%		\begin{subfigure}{0.5\textwidth}
%		\centering
%		%\includegraphics[width= 1 \textwidth]{GaAsnanoDBR2.eps}
%		\includegraphics[width= 1 \textwidth]{eff2nd1D.eps}
%		\caption{}
%		\label{eff2nd1D}
%	\end{subfigure}
%	\caption{(a) Efficiency for the 2nd antinode. (b) Efficiency for the 1st antinode.}
%	\label{effresult}
%\end{figure}
\begin{figure}[ht]
	%\advance\leftskip-4cm
	\begin{subfigure}{1\linewidth}
		\centering
		\includegraphics[width= 1 \linewidth]{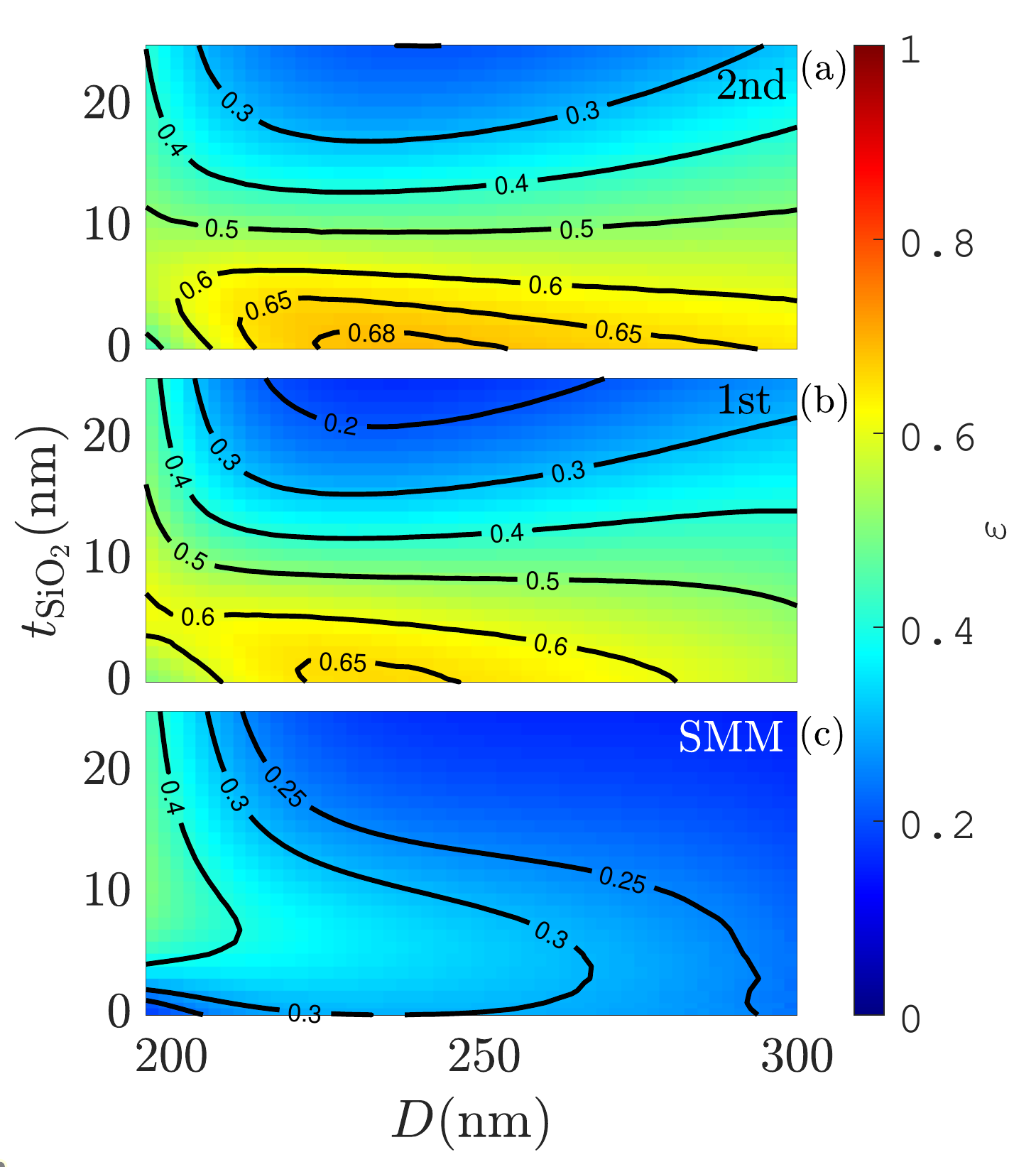}
	\end{subfigure}
	%\caption{(a) $\varepsilon$ for the 2nd antinode. (b) $\varepsilon$ for the 1st antinode. (c) $\varepsilon$ using the 1D model. $NA=0.75$}
	\caption{Collection efficiency, $\varepsilon$ ($NA=0.75$), for a QD at the 2nd antinode (a), 1st antinode (b) and any antinode using the SMM (c) as a function of the diameter, $D$, and the silica layer thickness, $t_{\rm SiO_2}$.}
	\label{eff_result}
\end{figure}

The source collection efficiency, $\varepsilon$, for the 2nd antinode, the 1st antinode and the SMM are shown in Fig.\ (\ref{eff_result}) for a numerical aperture of $\mathrm{NA}=0.75$. Here the overall performance for the 2nd antinode is slightly better than for the 1st antinode, and a maximum value of $\varepsilon=0.69$ is reached at $D=\SI{238}{nm}$ and $t_{\rm SiO_2}=\SI{0}{nm}$. Despite the very simple geometry of the nanopost, a surprisingly high collection efficiency of $\varepsilon=0.69$ is achievable. In general, it can be ascertained that no silica layer is more beneficial for the efficiency, which is surprising compared to the traditional photonic nanowire, where the silica layer enhances the modal reflectivity \cite{Claudon2010} and thus the collection efficiency.
Now comparing the efficiency of the SMM to the full model, there are substantial differences. This is unlike the micropillar and the photonic nanowire geometries for which the single-mode model $\varepsilon _{\rm s} = \beta \gamma$ is an excellent \cite{Friedler2009,Gregersen2016,Wang2020_PRB_Biying} approximation. For the nanopost, the SMM efficiency is much smaller in the entire parameter space except for the smallest diameters. This clearly shows that there are important physical mechanisms which are not accounted for in the SMM. Furthermore, comparing the figures for the Purcell factors and the efficiencies, there seems to be no apparent correlation between the two. This is also a surprising result compared to traditional Fabry-Pérot cavities and indicates that there are different physical mechanisms at play which govern the Purcell factor and the efficiency. 

In Supplementary \ref{sec:NA}, we vary the numerical aperture and present its influence on the collection efficiency. In Supplementary \ref{sec:Gauss}, we present the collection efficiency taking into consideration the overlap with a Gaussian profile.

\section{Theory} \label{Sec:Theory}

\subsection{Method}

We use an eigenmode method combined with a standard scattering matrix formalism \cite{NUMERICALMETHOD2014} in the frequency domain. In this method, the structure is divided into layers of uniform permittivity along the propagation direction. In each layer, the electrical field is expanded on the eigenmodes, and the scattering matrices are used to connect the eigenmodes at the interfaces between the layers. The eigenmodes are obtained using the Fourier modal method (FMM) with open boundary conditions \cite{Gur2021a}, which provides direct access to the modes needed to understand the physics.
The nanopost structure is split into four layers: the bottom gold substrate, the silica layer, the nanowire and the top air region. Due to the cylindrical symmetry of the nanowire, a cylindrical coordinate system is used. 
The QD is modelled as a classical point dipole, and we make use of the relationship $\Gamma/\Gamma_{\rm Bulk}=P/P_{\rm Bulk}$ to calculate the Purcell factor \cite{novotny2012principles}. $P$ is the emitted power of the dipole, and $P_{\rm Bulk}$ is the power emitted in a bulk medium, and thus the Purcell factor is defined as $F_{\rm p}=P_{\rm T}/P_{\rm Bulk}$. The second quantity of interest is the collection efficiency defined as $\varepsilon=P_{\rm collected}/P_{\rm T}$, where $P_{\rm collected}$ is the power collected in the far-field of a lens with numerical aperture $\rm NA$.

\subsubsection{Eigenmodes}

The electrical field for an eigenmode has the following expression:
\begin{equation}
	\textbf{E}_j(r,\phi,z)=\textbf{e}_j(r,\phi)\exp{(i\beta_j z)},
\end{equation} where $j$ refers to eigenmode index, $\textbf{e}_j(r,\phi)$ is the mode profile for the given eigenmode and $\beta_j$ is the propagation constant. For three out of the four layers, the permittivity profile is constant over the entire plane. In this case, the eigenmodes are simply cylindrical plane waves. These eigenmodes exist as a continuum where the propagation constant takes the value $\beta=\sqrt{(n_{\rm layer}k_0)^2-k_{\perp}^2}$, where $n_{\rm layer}$ is the refractive index, $k_0$ is the free-space wavenumber and $k_{\perp}$ is the in-plane $k$-value. $k_{\perp}$ can take any value $k_{\perp}\in[0,\infty]$ and eigenmodes exist for all the values. For each $\beta$, two orthogonal solutions exist, which can be separated into TE modes ($e_z=0$) and TM modes ($h_z=0$). For layers with real-valued refractive index, this continuum can be separated into radiation modes satisfying $0<(\beta)^2\leq(n_{\rm layer}k_0)^2$ and evanescent modes $(\beta)^2<0$ which decays exponentially and carry no power according to the Poynting vector. Specifically for the air layer, the propagation constant of the radiation modes can directly be interpreted as the propagation angle with respect to the z-axis using the expression $\theta=\arccos{(\beta/k_0)}$. Expanding a forward propagating electrical field on the eigenmodes will have the following expression:
\begin{equation}
	\textbf{E}(r,\phi,z)=\sum_{s=1}^2\int_0^{\infty}a_s(k_{\perp})\textbf{e}_s(r,\phi,k_{\perp})\exp{(i\beta(k_{\perp}) z)}d k_{\perp},
\end{equation} where $s$ refers to the two solutions and $a_s(k_{\perp})$ is the expansion coefficient. Numerically, the continuum is discretized into $N$ modes and truncated with a cut-off value for $k_{\perp}$ \cite{Hayrynen2016,Gur2021a}, which will lead to the following expression:
\begin{equation}
	\textbf{E}(r,\phi,z)=\sum_{j=1}^{N}a_j\textbf{e}_j(r,\phi)\exp{(i\beta_j z)},
\end{equation} where $s$ is absorbed into $j$, and the $\Delta k_{\perp}$ that would appear due to the discretization is absorbed into definition of the eigenmode profile.

%For the nanowire there is an additional class of eigenmodes, namely the guided modes. The propagation constant of the guided modes satisfy the condition $(n_{GaAs}k_0)^2>(\beta_j)^2>(n_{\rm air}k_0)^2$ and the description of this class of modes can be found in many textbooks.

The first class of eigenmodes for the nanowire are the guided modes for which the propagation constants satisfy the condition $(n_{\rm air}k_0)^2<(\beta_j)^2<(n_{\rm GaAs}k_0)^2$. The guided mode is confined to the core of the nanowire, and outside the nanowire, the field decays. There is a finite number of guided modes, and there will always be at least one guided mode, the fundamental $\mathrm{HE}_{11}$ mode. The description of guided modes can be found in various textbooks \cite{Yariv_6th}. The second class of eigenmodes is the background continuum, very similar to the continuum of the air layer. These eigenmodes can be viewed as perturbed versions of the cylindrical plane waves of the air layer and be separated into radiation and evanescent modes the exact same way. There are also two orthogonal solutions, but these can no longer be separated into pure TE and TM modes. Studies of this class of eigenmodes are plentiful in the literature \cite{Snyder1971,SammutPHD,Nyquist1981,Vassallo81,Sammut1982,Vassallo83,Snyder1983,Tigelis1987,Morita1988,Alvarez2005}.% Naturally, a single radiation mode (or a very small part of the continuum) of the nanowire will experiences a small reflection at the top interface between the nanowire and air and will mainly transmit into the eigenmodes with the same propagation constant. However, later on we will show that...     

\subsubsection{Dipole emission in an infinite structure}

The QD is modelled as a classical point dipole with in-plane orientation and harmonic time dependence at the frequency $\omega$. The corresponding current density is $\mathbf{J}(\mathbf{r})=-i\omega\mathbf{p}\delta(\mathbf{r}-\mathbf{r}_{\rm d})$, where $\mathbf{r}_{\rm d}$ is the position of the QD and $\mathbf{p}$ is the dipole moment. The emitted power of the dipole can be calculated as \cite{novotny2012principles}:
\begin{equation}
	\begin{split}
		P = - \frac{1}{2}\int_V{\rm Re}[{{\bf J}^*({\bf r}_{\rm})\cdot{\bf E}(\bf r_{\rm})}]dV= \frac{\omega}{2} {\rm Im}[{{\bf p} \cdot{\bf E}(\bf r_{\rm d})}] .    
	\end{split}
	\label{Power}
\end{equation} The total emitted power, $P_{\rm T}$, can be calculated by evaluating the total field, $\mathbf{E}_{\rm T}(r_{\rm d})$, but also the power into individual modes, $P_{j}$, by evaluating $\mathbf{E}_{j}(r_{\rm d})$. Placing the dipole on-axis inside an infinitely long nanowire with position $z_{\rm J}$ will result in the following electrical field:\begin{equation}
	\begin{split}
		\mathbf{E}_{\rm T}(\mathbf{r})=\sum_{j}a_{j}^{\rm J}\mathbf{e}_{j}^{+}(\mathbf{r}_{\perp})\exp(i\beta_{j} (z-z_{\rm J})) \ (z>z_{\rm J})
		\label{Etotal}
	\end{split}
\end{equation} \begin{equation}
	\begin{split}
		\mathbf{E}_{\rm T}(\mathbf{r})=\sum_{j}b_{j}^{\rm J}\mathbf{e}_{j}^{-}(\mathbf{r}_{\perp})\exp(-i\beta_{j} (z-z_{\rm J})) \ (z<z_{\rm J}),
		\label{Etotal_back}
	\end{split}
\end{equation} where the superscript $+$ refers to forward propagating and $-$ refers to backward propagating. For the electrical fields, the forward and backwards propagating fields are identical:  $\mathbf{e}_{j}=\mathbf{e}_{j}^{+}=\mathbf{e}_{j}^{-}$. $a_{j}^{\rm J}$ and $b_{j}^{\rm J}$ are the field expansion coefficients and in an infinite structure, such as the nanowire, have the following simple expressions \cite{NUMERICALMETHOD2014}:
\begin{equation}
	\begin{split}
		a_{j}^{\rm J}=-\frac{-i\omega\mathbf{p}\cdot\mathbf{e}_{j}^{+}(\bf r_{\rm d})}{2}
		\label{aj}
	\end{split}
\end{equation}
\begin{equation}
	\begin{split}
		b_{j}^{\rm J}=-\frac{-i\omega\mathbf{p}\cdot\mathbf{e}_{j}^{-}(\bf r_{\rm d})}{2}.
		\label{bj}
	\end{split}
\end{equation} The expansion coefficients can then be represented by vectors: 
\begin{equation}
	\begin{split}
		\mathbf{a}_{\infty \rm NW}=\begin{bmatrix}
			a_{1}^{\rm J} & a_{2}^{\rm J} & \cdots & a_{N}^{\rm J}
		\end{bmatrix}
		\label{ajvector}
	\end{split}
\end{equation}
\begin{equation}
	\begin{split}
		\mathbf{b}_{\infty \rm NW}=\begin{bmatrix}
			b_{1}^{\rm J} & b_{2}^{\rm J} & \cdots & b_{N}^{\rm J}
		\end{bmatrix},
		\label{bjvector}
	\end{split}
\end{equation} where subscript $\infty \rm NW$ refers to the infinite nanowire.

%$\mathbf{a}_{\infty NW}=\begin{bmatrix}
	%    a_{1}^{\rm J} & a_{2}^{\rm J} & \cdots & a_{N}^{\rm J}
	%\end{bmatrix}$ and $\mathbf{b}_{\infty NW}=\begin{bmatrix}
	%    b_{1}^{\rm J} & b_{2}^{\rm J} & \cdots & b_{N}^{\rm J}
	%\end{bmatrix}$.
	
%	Equations similar to Eqs. (\ref{Etotal},\ref{Etotal_back}) can be derived for the magnetic field using the same expansion coefficients.

	\subsubsection{Multilayered structures and scattering matrices}
	
	%The reflection and transmission matrices are used to connect the field at the interfaces which is sketched in Fig.\ (\ref{Sketch2}). The coefficients in the matrices describes how much of a given mode is transmitted or reflected into another mode. A reflection matrix is shown in Eq. (\ref{refmatrix}).
	
	The reflection and transmission matrices are used to connect the field at the interfaces between the layers. These matrices are derived from the boundary condition that the tangential components of the electric and magnetic field are continuous across an interface \cite{NUMERICALMETHOD2014}. The coefficients in the matrices describe how a given mode is transmitted or reflected into another mode. A reflection matrix is shown in Eq. (\ref{refmatrix}).
	\begin{align}
	\begin{split}
	 		\mathbf{R}&=
		\begin{bmatrix}
			r_{11} & r_{12} & r_{13} & \dots  & r_{1N} \\
			r_{21} & r_{22} & r_{23} & \dots  & r_{2N} \\
			\vdots & \vdots & \vdots & \ddots & \vdots \\
			r_{N1} & r_{N2} & r_{N3} & \dots  & r_{NN}
		\end{bmatrix}.\\    
		\label{refmatrix}   
	\end{split}
	\end{align}	The $r_{11}$ coefficient represents the reflection of the fundamental mode back into itself, while the remaining part of the first column represents the reflection of the fundamental mode into all other modes and so forth. Propagation matrices are used to propagate the field inside a layer and are defined in the following way:
	\begin{align}
	\begin{split}
		\mathbf{P}(z)&=
		\begin{bmatrix}
			e^{ i \beta_{1} z} & 0 & 0 & \dots  & 0 \\
			0 & e^{ i \beta_{2} z} & 0 & \dots  & 0 \\
			\vdots & \vdots & \vdots & \ddots & \vdots \\
			0 & 0 & 0 & \dots  & e^{ i \beta_{N} z}
		\end{bmatrix}.\\    
	\end{split}
	\end{align} 	The total field inside the structure can then be calculated by taking into account the round-trips which the initially emitted light takes inside the cavity. Above the emitter, the field takes the following expression \cite{NUMERICALMETHOD2014}:
	\begin{equation}
		\begin{split}
			\mathbf{E}_{\rm T}(\mathbf{r})&=\sum_{j}a_{\mathrm{tot},j}^{\rm J}\mathbf{e}_{j}^{+}(\mathbf{r}_{\perp})\exp(i\beta_{j} (z-z_{\rm J}))\\
			&+b_{\mathrm{tot},j}\mathbf{e}_{j}^{-}(\mathbf{r}_{\perp})\exp(-i\beta_{j} (z-z_{\rm J}))\quad (z>z_{\rm J}),
			\label{eigEdE}
		\end{split}
	\end{equation}	where the new expansion coefficients are calculated using the following equation \cite{NUMERICALMETHOD2014}:
	\begin{equation}
		\begin{split}
			\mathbf{a}_{\rm tot}^{\rm J}&=(\mathbf{I}-\mathbf{P}(h_{\rm b})\mathbf{R}_{\rm bot}\mathbf{P}(h_{\rm b})\mathbf{P}(h_{\rm t})\mathbf{R}_{\rm top}\mathbf{P}(h_{\rm t}))^{-1}\\
			&(\mathbf{a}_{\infty \rm NW}+\mathbf{P}(h_{\rm b})\mathbf{R}_{\rm bot}\mathbf{P}(h_{\rm b})\mathbf{b}_{\infty \rm NW})
			\label{atot}
		\end{split}
	\end{equation}	and
	\begin{equation}
		\begin{split}
			\mathbf{b}_{\rm tot}=\mathbf{P}(h_{\rm t})\mathbf{R}_{\rm top}\mathbf{P}(h_{\rm t})\mathbf{a}_{\rm tot}^{\rm J}.
			\label{btot}
		\end{split}
	\end{equation}	Now the Purcell factor can be calculated by evaluating the total field at the dipole position using Eq. (\ref{Power}). 
	
	\subsubsection{Far-field and efficiency}
	
	To obtain the field in the air above the structure, we apply the propagation and transmission matrix on the forward propagating light in the cavity and thus obtain the expansion coefficients.
	\begin{equation}
		\begin{split}
			\mathbf{a}_{\rm air}=\mathbf{T}_{\rm top}\mathbf{P}(h_{\rm t})\mathbf{a}_{\rm tot}^{\rm J}.
			\label{aair}
		\end{split}
	\end{equation}	To calculate the collected power in a lens with some numerical aperture, a near- to far-field transformation is used \cite{Balanis89}. The far-fields $\mathbf{E}_{\textrm{FF}}(R,\theta,\phi)$ and $\mathbf{H}_{\textrm{FF}}(R,\theta,\phi)$  are calculated on the surface of a sphere with radius R, and the radial component of the resulting Poynting vector is:
	\begin{equation}
		\begin{split}
			S_{\textrm{FF}}(R,\theta,\phi)=(E^*_{\textrm{FF},\theta}H_{\textrm{FF},\phi}-E^*_{\textrm{FF},\phi}H_{\textrm{FF},\theta}).
			\label{Spoynt}
		\end{split}
	\end{equation} The collected power in the far-field is then:
	\begin{equation}
		\begin{split}
			P_{\textrm{FF}}(\textrm{NA})&=\frac{1}{2}R^2\int_0^{2\pi}\int_0^{\theta_{\textrm{NA}}}S_{\textrm{FF}}(R,\theta,\phi)\sin(\theta)d\theta d\phi\\
			&=\int_0^{2\pi}\int_0^{\theta_{\textrm{NA}}}p_{\textrm{FF}}(\theta,\phi)\sin(\theta)d\theta d\phi,\\
			\label{PowerFF}
		\end{split}
	\end{equation} where $p_{\textrm{FF}}(\theta,\phi)$ is the power per unit solid angle in the far-field and $\theta_{\textrm{NA}}$ is determined by the $\mathrm{NA}$ ($\textrm{NA}=\sin(\theta_{\textrm{NA}})$). The $R$ dependence cancels out as the Poynting vector scales as $1/R^2$. %Later on $p_{\textrm{FF}}(\theta,\phi)$ will be used to plot the far-fields.

	\subsubsection{Single-mode Fabry-Pérot model}
	
	When calculating the Purcell factor using the SMM, we only consider the fundamental mode. The SMM equations equivalent to Eq. (\ref{atot}-\ref{btot}) are
	
		\begin{equation}
		\begin{split}
			\mathrm{a}_{\rm tot,SMM}^{\rm J}&=a_1^{\rm J}\frac{1+r_{11,\rm bot}\mathrm{e}^{i2h_{\rm b}\beta_1}}{1-r_{11,\rm bot}r_{11,\rm top}\mathrm{e}^{i2h_{\rm total}\beta_1}}
			\label{1Da}
		\end{split},
	\end{equation} and
	\begin{equation}
		\begin{split}
			\mathrm{b}_{\rm tot,SMM}&=\mathrm{a}_{\rm tot,SMM}^{\rm J}r_{11,\rm top}\mathrm{e}^{i2h_{\rm t}\beta_1}
			\label{1Db}
		\end{split}.
	\end{equation} The collected power in the far-field is then calculated by inserting Eq. (\ref{1Da}) into Eq. (\ref{aair}), and the SMM efficiency is then defined as $\varepsilon_s=P_{\rm collected,SMM}/P_{\rm T}$, equivalent to the definition in the introduction.

	\subsection{Mode-coupling and emission channels}

\begin{figure}[h]
		%\advance\leftskip-4cm
		\begin{subfigure}{1\linewidth}
			\centering
			\includegraphics[width= 1 \linewidth]{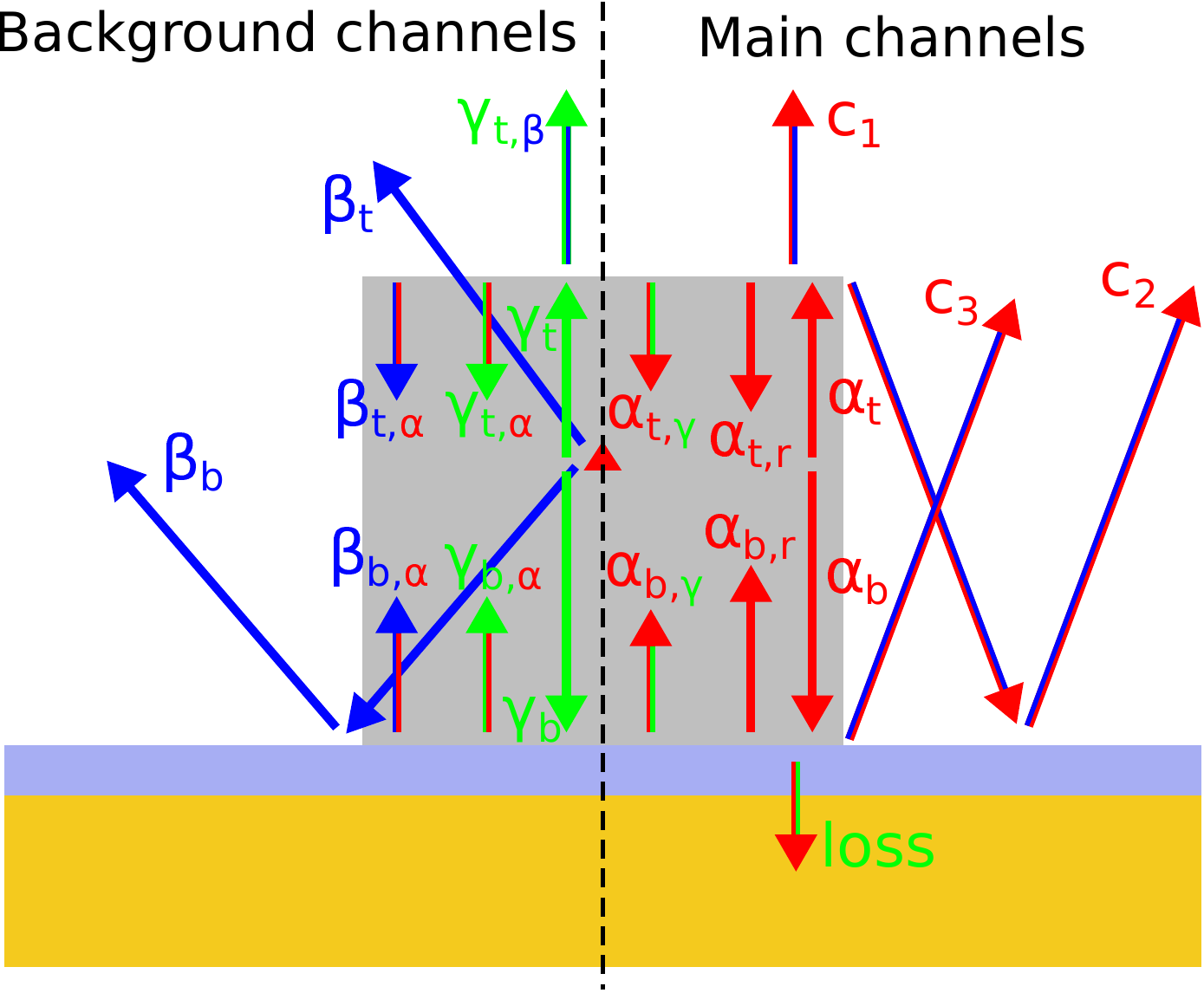}
		\end{subfigure}
		\caption{Sketch of the nanopost and the different emission channels and examples of mode-coupling. The emission channels are separated into the main channels and the background channels. The red color corresponds to the fundamental mode, the blue color corresponds to radiation modes and the green color corresponds to evanescent modes. Some arrows have two colors, where the color of the arrowhead corresponds to the original channel, but the 2nd color on the shaft signifies the current mode classification. As an example consider $c_1$ which is the transmission of the fundamental mode into the air. It has a partly blue shaft as it is now classified as radiation, but it originated as the fundamental mode (red arrowhead). See main text for more information.}
		\label{channels}
	\end{figure}

	Important coupling effects take place at the top and bottom interfaces of the nanowire. At both interfaces, all the modes couple to each other, i.e. all the elements in the reflection matrices are non-zero; however, some modes and elements are more important than others. In Fig.\ (\ref{channels}), different examples of mode coupling are shown along with the emission channels that will contribute to the far-field. The sketch is divided into two parts: the main channels and the background channels. The main channels consist of all the light that originated as the fundamental mode, $\alpha_{\rm t}$ and $\alpha_{\rm b}$, which is indicated by the red arrowheads. This is the propagating mode that experiences sufficiently large reflections at both interfaces such that it is Purcell enhanced.
	
	\subsubsection{Path of the fundamental mode}
	
	Let us now follow the path of the fundamental mode. The light emitted into the fundamental mode will propagate upwards and downwards indicated by $\alpha_{\rm t}$ and $\alpha_{\rm b}$. At the top interface the fundamental mode is:
	\begin{itemize}[noitemsep]
	    \item Transmitted into the air indicated by $c_1$.
	    \item Reflected, indicated by $\alpha_{\rm t,r}$  (the channel responsible for Purcell enhancement).
	    \item Scattered into radiation that propagates downwards indicated by the red/blue arrow pointing towards the bottom mirror. This radiation will then be reflected by the bottom mirror and then be transmitted into the air indicated by $c_2$.
	    \item Coupled to evanescent modes indicated by $\alpha_{\rm t,\gamma}$.
	\end{itemize}	At the bottom interface the fundamental mode is:
		\begin{itemize}[noitemsep]
	    \item Transmitted into the mirror and lost indicated by the green/red arrow pointing downwards at the very bottom.
	    \item Reflected, indicated by $\alpha_{\rm b,r}$ (the channel responsible for Purcell enhancement).
	    \item Scattered into radiation that propagates towards the air indicated by $c_3$.
	    \item Coupled to evanescent modes indicated by $\alpha_{\rm b,\gamma}$.
	\end{itemize}

	%\begin{itemize}
	%    \item Transmitted into the air indicated by $c_1$.
	%    \item Reflected indicated by $\alpha_{t,r}$  (the channel responsible for Purcell enhancement).
	%\end{itemize}
	
	%\begin{enumerate}
	%    \item 
	%\end{enumerate}
	
	As we will demonstrate in the following, the three radiation channels $c_1$, $c_2$ and $c_3$ are the main channels that will contribute to the far-field.
	
	%One process that is not shown in Fig.\ (\ref{channels}) is the back-coupling of the initially scattered radiation, i.e. $\textrm{HE}_{11}\rightarrow\textrm{radiation/evanescent}\rightarrow\textrm{HE}_{11}$. This process is a recycling effect which is important for the Purcell factor and later on we will present models which can quantify the contributions of the different channels and mode coupling, including this recycling effect.
	
	\subsubsection{Path of the background emission}
	%In general the lower the Purcell factor, the more important these channels are for $P_{\rm T}$ and $P_{\rm collected}$.
	Let us now consider the background emission channels. First, we have the light directly emitted into radiation, indicated by the blue arrows of $\beta_{\rm t}$ and $\beta_{\rm b}$. The radiation can both be emitted upwards or downwards and then reflected by the mirror. At both interfaces, a small part of the radiation modes can also couple to the fundamental mode indicated by the blue/red arrows of $\beta_{\rm t,\alpha}$ and $\beta_{\rm b,\alpha}$. We also have light coupled to the evanescent modes, which is indicated by the long green arrows pointing upwards and downwards of $\gamma_{\rm t}$ and $\gamma_{\rm b}$. These modes do not propagate in a traditional sense, but at the interfaces, they can scatter into the fundamental mode indicated by the green/red arrows of $\gamma_{\rm t,\alpha}$ and $\gamma_{\rm b,\alpha}$ at the top and bottom interfaces. At the top interface, the evanescent modes can also couple to radiation and be transmitted indicated by the green/blue arrow pointing upwards, $\gamma_{\rm t,\beta}$.

\section{Analysis of the Purcell factor and efficiency} \label{Sec:Analysis}

\begin{figure}[b]
	%\centering
	%\advance\leftskip-4cm
\begin{subfigure}{1\linewidth}
		\centering
		\includegraphics[width= 1 \linewidth]{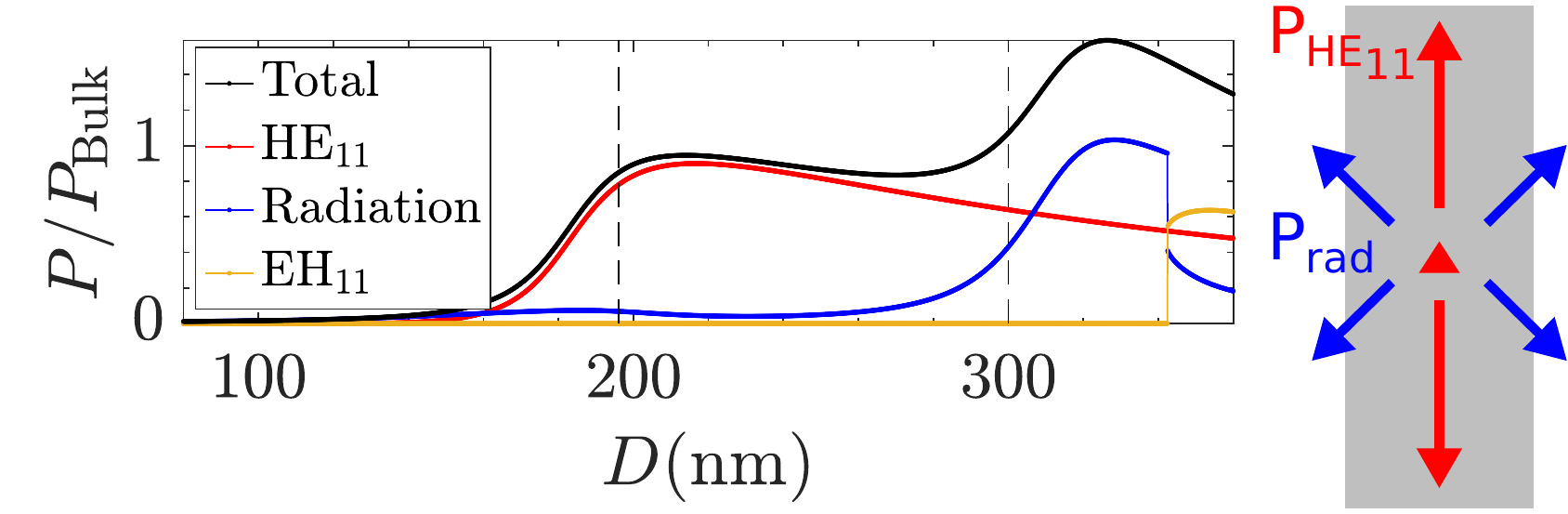}
		%		  \makebox[\textwidth][c]{\includegraphics[width=1\textwidth]{power_nanowirev3.eps}}%

	\end{subfigure}
	\caption{Power emission in the infinite nanowire as a function of the diameter, $D$. A sketch of the emission in an infinite nanowire with an embedded QD is shown in the right part of the figure.}
	\label{power_nanowire}
\end{figure}

The starting point of the analysis is the emission rates in the infinitely long GaAs nanowire, which directly represent the initial coefficients of Eq. (\ref{ajvector}-\ref{bjvector}) through Eq. (\ref{Power}-\ref{Etotal_back}).
In Fig.\ (\ref{power_nanowire}), the emission rates for the present guided modes, the radiation modes and the total emission are shown as a function of the nanowire diameter. The black dotted vertical lines represent the interval which is used in the full simulations of the nanopost. In this interval, the infinite nanowire only contains one guided mode, the fundamental $\mathrm{HE}_{11}$ mode, and most of the power is emitted into this mode. The emission into radiation is suppressed in most of the interval and only begins to increase when the diameter reaches $\SI{300}{nm}$ before the $\mathrm{EH}_{11}$ mode appears. The emission rates thus show that the radiation background channel (the initial coefficients for the radiation) presumably only has a minor influence on the Purcell factor and efficiency as long as the cavity does not suppress the fundamental mode.
Fig.\ (\ref{power_nanowire}) cannot be used to quantify the importance of the evanescent background channels, as the field components of the evanescent eigenmodes only have a real part. Later on, in the third subsection, models will be used to quantify the effect of the evanescent background channels the finite-length nanopost structure.

\subsection{Scattering of the fundamental mode at the interfaces}

Here, we will study reflection and transmission at the top interface between the GaAs nanowire and the air above. The fundamental mode is launched towards the interface, and then the reflection and transmission coefficients are calculated. The power reflection coefficient for reflection of mode $n$ into mode $m$ is $R_{m,n} = |r_{m,n}|^2$, and the power transmission coefficients are defined similarly. The total power reflection of the fundamental into radiation is then defined as $R_{\mathrm{rad},1}=\sum_{s=1}^{2}\int_0^{k_0}|r_{s,1}(k_{\perp})|^2d k_{\perp}$, and in the discretized regime $R_{\mathrm{rad},1}=\sum_{n=2}^{N_{k_0,\mathrm{wire}}+1}|r_{n,1}|^2$, where the index $N_{k_0,\mathrm{wire}}$ corresponds to the total number of radiation modes in the nanowire. Thus the total power reflection of the fundamental mode is $R_{\mathrm{total},1}=R_{1,1}+R_{\mathrm{rad},1}$. The total power transmission of the fundamental mode is then defined as $T_{\mathrm{total},1}=\sum_{n=1}^{N_{k_0,\mathrm{air}}}|t_{n,1}|^2$ and due to power conservation we have $R_{\mathrm{total},1}+T_{\mathrm{total},1}=1$.

%Note: It can be difficult to quantify reflection into evanescent modes as these modes do not carry any power according to the Poynting vector. 

%1 or 2 Figures plotting $R_{11}$, $T$, $R_{rad}$ and $R_{\rm T}$ as a function of $D/\lambda$.

\begin{figure}[h]
	%\advance\leftskip-4cm
	\begin{subfigure}{1\linewidth}
		\centering
		\includegraphics[width= 1 \linewidth]{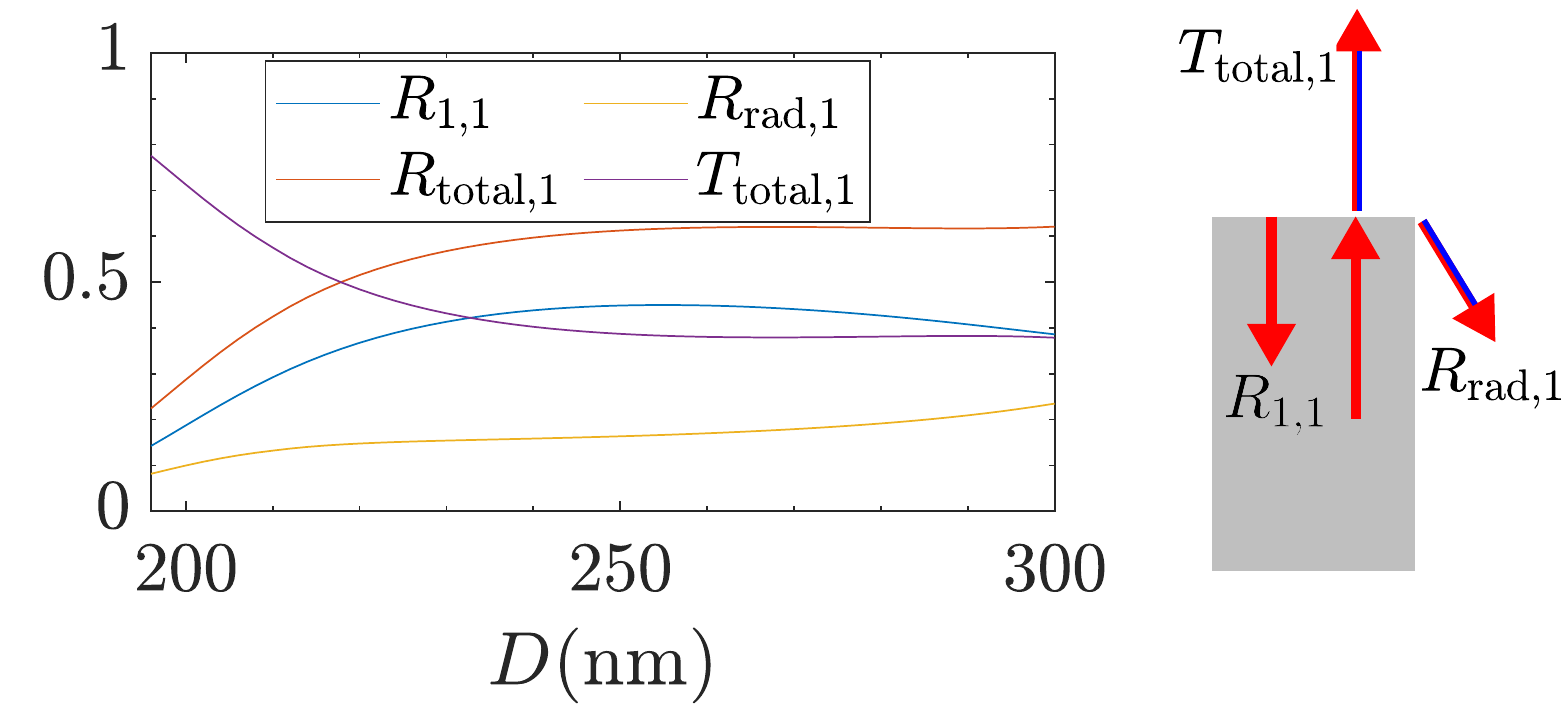}
		%  \makebox[\textwidth][c]{\includegraphics[width=1\textwidth]{topreflectionv3.eps}}%

	\end{subfigure}
	\caption{Reflection and transmission of the fundamental mode as a function of the diameter, $D$, at the top interface. A sketch of top interface and the reflection and transmission of the fundamental mode is seen to the right.}
	\label{refHE11top}
	
\end{figure}

In Fig.(\ref{refHE11top}), the power reflections, along with the power transmission of the fundamental mode, are shown as a function of the nanowire diameter $D$. By comparing the magnitudes of the modal reflection ($R_{1,1}$) and the reflection into radiation ($R_{\mathrm{rad},1}$), it is clear that the reflection into radiation ($c_2$ in Fig.(\ref{channels})) is essential for the far-field. This indicates why the SMM fails to describe the efficiency. This mechanism will have a much smaller influence on the Purcell factor as the radiation modes only has a small field amplitude at the center of the nanowire shown in the previous section (Fig.\ (\ref{power_nanowire})). However, the reflection matrix is approximately symmetric, such that the first column and the first row is identical, $r_{1,n}\approx r_{n,1}$. This means that a small part of the radiation will actually scatter back into the fundamental mode at the top interface, and this will have an influence on the Purcell factor, which we will show later. In general for Fig.\ (\ref{refHE11top}), we observe small modal reflections for small diameters and this will lead to a limited cavity effect and thus a lower Purcell factor.

\begin{figure}[htb]
	%\advance\leftskip-4cm
	\begin{subfigure}{1\linewidth}
		\centering
		\includegraphics[width= 1 \linewidth]{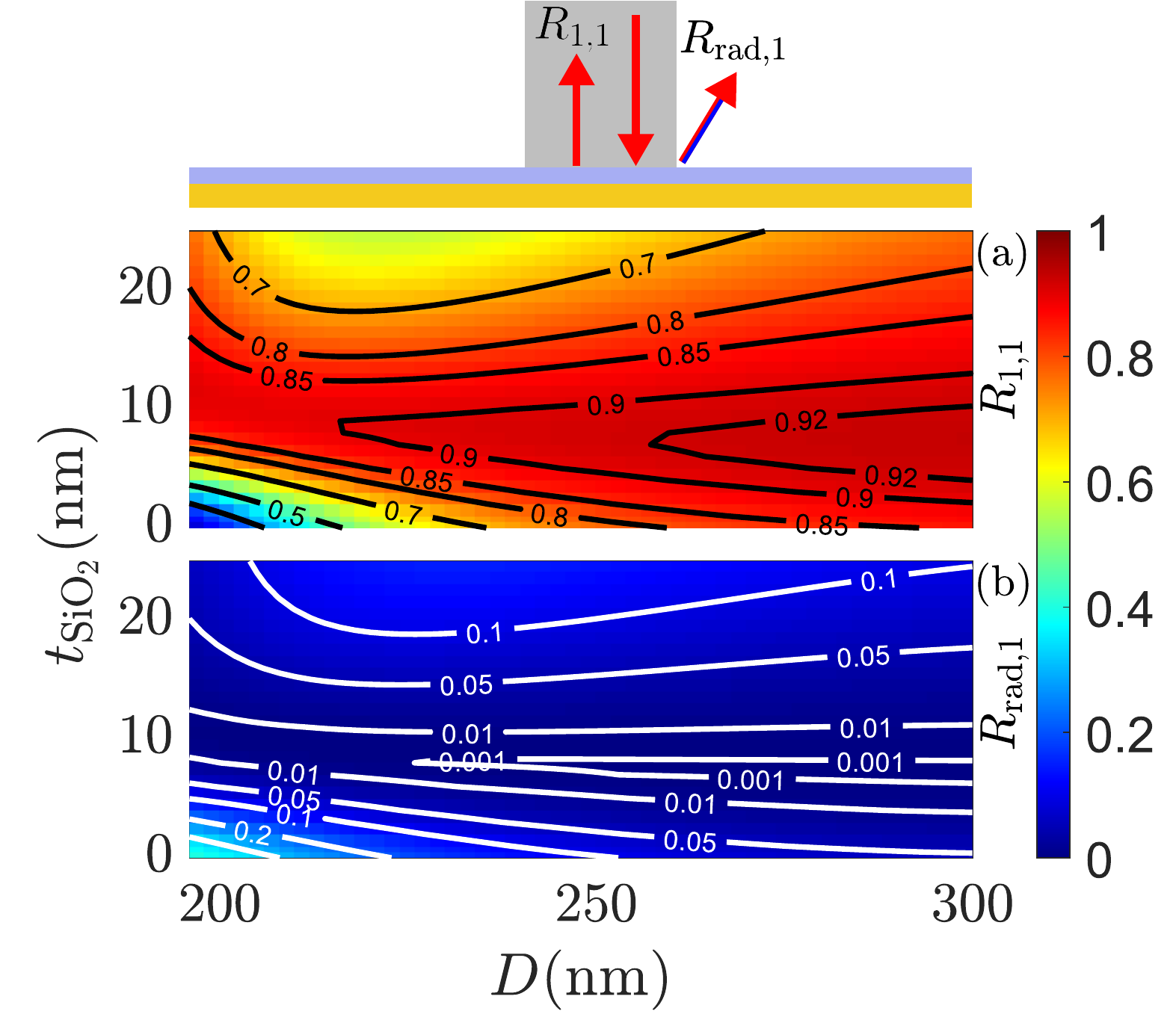}
	\end{subfigure}
	\caption{(a) Modal reflection, $R_{1,1}$ and (b) reflection into radiation of the fundamental mode, $R_{rad,1}$, as a function of the diameter, $D$, and silica layer thickness, $t_{\mathrm{SiO_2}}$, at the bottom interface. A sketch of the bottom interface is shown in the top part of the figure.}
	\label{refbotfig}
\end{figure}

We now consider the reflections at the bottom interface between the GaAs nanowire and the silica-gold mirror. Compared to the top interface, there is now an additional parameter, namely the thickness of the silica layer, $t_{\rm SiO_2}$. The purpose of the silica layer is to increase the reflection of the fundamental mode and avoid coupling to surface plasmons which would decrease the reflection \cite{Friedler:08}.   
In Fig.\ (\ref{refbotfig}), the bottom reflection coefficients are shown as a function of the diameter and the silica layer thickness. In the parameter ranges where the modal reflection is large $\sim0.9$ (strong cavity effect), the reflection into radiation is small $\sim0.01$. Here we do not expect a significant contribution of the scattering into radiation, $c_3$. However, as the modal reflection decreases, the reflection into radiation increases to larger values $\sim0.1$, and here $c_3$ will contribute to the far-field. For small diameters and low values of the silica layer thickness, the modal reflection is small (weak cavity effect) and the scattering into radiation is very large $\sim0.2$. This behavior has been described in the literature\cite{Friedler:08}.

%Here we will study reflections at the bottom interface between the GaAs nanowire and the silica-metal mirror. In the first simulation, the fundamental mode is launched towards the interface and then we will calculate the reflection of the fundamental mode, $R_{11}$, the reflection into radiation modes, $R_{rad}$ and the total reflection, $R_{\rm T}=R_{11}+R_{rad}$. Here we have two free parameters to vary, $D/\lambda$ and $t_{SiO_2}/\lambda$. 

%1 or 2 Figures plotting $R_{11}$, $R_{rad}$ and $R_{\rm T}$ as a function of $D/\lambda$ and $t_{SiO_2}/\lambda$.
%\begin{figure}[H]
%	\begin{subfigure}{0.5\textwidth}
%		\centering
%		%\includegraphics[width= 1 \textwidth]{GaAsnanoDBR2.eps}
%		\includegraphics[width= 1 \textwidth]{botreflection11v3.eps}
%		\caption{}
%		\label{refbot11}
%	\end{subfigure}
	%\begin{subfigure}{0.5\textwidth}
	%\centering
	%%\includegraphics[width= 1 \textwidth]{GaAsnanoDBR2.eps}
	%\includegraphics[width= 1 \textwidth]{botreflectiontotal.eps}
	%\label{refbottot}
	%\end{subfigure}
%	\begin{subfigure}{0.5\textwidth}
%		\centering
%		%\includegraphics[width= 1 \textwidth]{GaAsnanoDBR2.eps}
%		\includegraphics[width= 1 \textwidth]{botreflection1radv3.eps}
%		\caption{}
%		\label{refbotrad}
%	\end{subfigure}
%	\caption{(a) Modal reflection and (b) reflection into radiation of the fundamental mode as a function of the diameter and silica layer thickness.}
%	\label{refbotfig}
%\end{figure}

%In Fig.\ (\ref{refbotfig}) the bottom reflections are shown as a function of the diameter and the silica layer thickness. Over a large parameter range the modal reflection is above $0.9$, which leads to a good cavity effect. In this region, the scattering into radiation is very small 

%For small diameters and low values of the silica layer thickness, the modal reflection is small and the cavity effect will be weak. Here the reflection into radiation is larger and above $0.1$. This behavior has been described in the literature\cite{Friedler:08}. For increased values of the diameter or the silica layer thickness, the modal reflection increases and over a large range of parameters, the modal reflection is above $0.8$ and even above $0.9$, which leads to a good cavity effect. In this regime, the reflection into radiation is below $0.05$ and decreases further as the modal reflection increases. Therefore the reflection into radiation at the top interface has a larger contribution compared to the reflection at the bottom interface in this region.  

Due to the scale invariance of Maxwell's equations, both the top and bottom reflections are broadband, which gives the potential for broadband Purcell enhancement.

\subsection{Enhanced efficiency}

In the first part of this subsection, we present different methods to model the efficiency to show how important the different emission channels are. Then we will apply the modelling methods on the structure with the largest efficiency, namely $D=\SI{238}{nm}$ and $t_{\rm SiO_2}=\SI{0}{nm}$ with a QD placed at the 2nd antinode.

%In the first part of this subsection we present different methods to model the efficiency and the Purcell factor to investigate the contributions of the different channels. In the second part we will apply the modelling methods on two different structures from the parameter scan. One structure provides a good collection efficiency and the other a good Purcell factor. To ease the comparison the same diameter, $D=\SI{246}{nm}$ is chosen, but two different silica thicknesses, $t_{SiO_2}=\SI{0}{nm}$ and $t_{SiO_2}=\SI{12}{nm}$, which we will refer to as structure A and structure B respectively.

%In the first part of this subsection we present different methods to model the efficiency and the Purcell factor to investigate the contributions of the different channels. In the second part we will apply the modelling methods on two different structures from the parameter scan. One structure provides a good collection efficiency and the other a good Purcell factor. To ease the comparison the same diameter, $D=\SI{246}{nm}$ is chosen, but two different silica thicknesses, $t_{SiO_2}=\SI{0}{nm}$ and $t_{SiO_2}=\SI{12}{nm}$, which we will refer to as structure A and structure B respectively.

%Maybe refer to structure A and B

%In this section, two specific designs of the nanopost choosen from the parameter scan will be studied in detail. Specifically, the contributions of the different channels to the efficiency and Purcell seen in Fig.(\ref{channels}) will be investigated. The first design has the parameters... 

\subsubsection{Efficiency contributions of the emission channels} \label{Modelling the efficiency}

We wish to separate and quantify the efficiency contributions of the main channels and the background channels shown in Fig.\ (\ref{channels}). We also wish to separate and quantify the direct emission of $c_1$ and the scattered channels of $c_2$ and $c_3$ shown in Fig.\ (\ref{channels}), and therefore we need two different methods. Recall that the efficiency is calculated as $\varepsilon=P_{\rm collected}/P_{\rm T}$. For both methods, $P_{\rm T}$ is calculated using the full model, but $P_{\rm collected}$ is calculated such that we can either 1. separate the main channels and the background channels or 2. separate the direct emission of $c_1$ and the scattered channels of $c_2$ and $c_3$.

In the first method, the reflection and transmission matrices are unchanged; however, a varying number of the initial coefficients, $\mathbf{a}_{\infty \rm NW}$ and $\mathbf{b}_{\infty \rm NW}$ of Eq. (\ref{ajvector}-\ref{bjvector}), are included when $\mathbf{a}_{\rm tot}^{\rm J}$ (Eq. (\ref{atot})) and thus $P_{\rm collected}$ is calculated. For instance, if we only include the first element of the initial coefficients, $a_{1}^{\rm J}$ and $b_{1}^{\rm J}$, and put the remaining elements to zero, then we only include the main channels in $P_{\rm collected}$, which originated as the fundamental mode. As we increase the number of initial coefficients included, the background channels are added starting from the first radiation mode until the last evanescent mode. $\varepsilon$ can then be plotted as a function of the included initial coefficients and if this curve is flat, then the main channels dominate the efficiency.

In the second method, the reflection and transmission matrices are also unchanged, all initial coefficients are included, but instead, a varying number of the final coefficients, $\mathbf{a}_{\rm tot}^{\rm J}$ of Eq. (\ref{atot}), are included when $P_{\rm collected}$ is calculated. The first element of $\mathbf{a}_{\rm tot}^{\rm J}$ represents all the light that ended up in the fundamental mode, where the main contribution is from the fundamental mode itself, but also includes contributions of the background channels which have scattered into the fundamental mode such as $\beta_{\rm t,\alpha}$, $\beta_{\rm b,\alpha}$, $\gamma_{\rm t,\alpha}$ and $\gamma_{\rm b,\alpha}$ seen in Fig.\ (\ref{channels}). However, by using the first method, we can quantify how strong these contributions are. If these contributions are weak, then the main channels are dominating. Thus if only the first element of the final coefficients is included in $P_{\rm collected}$, then only the direct transmission of the fundamental mode is included in the far-field, namely $c_1$ in Fig.(\ref{channels}). Then as we increase the number of elements of the final coefficients, the contributions of $c_2$ and $c_3$ are included. In principle, channels such as $\beta_{\rm t}$ and $\beta_{\rm b}$ are also included, but again we use the previous method to quantify these. $\varepsilon$ can then be plotted as a function of the included final coefficients, and if this curve is increasing, then the channels $c_2$ and $c_3$ are contributing to the efficiency.

Finally, we wish to visualize the interference between the direct transmission of the fundamental mode and the scattered channels by calculating the transmission of the fundamental mode and the entire background continuum separately: \begin{equation}
	\begin{split}
		\mathbf{a}_{\mathrm{air},\mathrm{HE}_{11}}=\mathbf{T}_{\rm top}\mathbf{P}(h_{\rm t})\mathbf{a}_{\mathrm{tot},\mathrm{HE}_{11}}^{\rm J}
		\label{f4f}
	\end{split}
\end{equation}
and\begin{equation}
	\begin{split}
		\mathbf{a}_{\rm air,BG}=\mathbf{T}_{\rm top}\mathbf{P}(h_{\rm t})\mathbf{a}_{\rm tot,BG}^{\rm J},
		\label{f4f}
	\end{split}
\end{equation} where
\def\A{
	\begin{bmatrix}
		a_{\rm tot,1} & 0 & \cdots & 0
\end{bmatrix}}
\begin{equation}
	\begin{split}
		\mathbf{a}_{\mathrm{tot},\mathrm{HE}_{11}}^{\rm J}=\A
		\label{f4f}
	\end{split}
\end{equation}
and
\def\B{
	\begin{bmatrix}
		0 & a_{\rm tot,2} & \cdots & a_{\mathrm{tot},N}
\end{bmatrix}}
\begin{equation}
	\begin{split}
		\mathbf{a}_{\rm tot,BG}^{\rm J}=\B.
		\label{f4f}
	\end{split}
\end{equation}
Then the phase difference between the two contributions can be calculated:\begin{equation}
	\begin{split}
		\Delta\phi=\arg(\mathbf{a}_{\mathrm{air},\mathrm{HE}_{11}})-\arg(\mathbf{a}_{\rm air,BG}).
		\label{f4f}
	\end{split}
\end{equation}The phase difference will be separated into TE and TM modes. Along with the phase difference, the far-field plots of the direct transmission of the fundamental mode ($\mathbf{a}_{\mathrm{air},\mathrm{HE}_{11}}$), the entire background continuum ($\mathbf{a}_{\mathrm{air},\mathrm{BG}}$) and the total field ($\mathbf{a}_{\rm air}$) will be shown.

\subsubsection{Influence of scattered radiation on collection efficiency}

%\paragraph{Efficiency.} 

%We will now apply the modelling methods. First, we study the efficiency of structure A with the parameters $D=\SI{246}{nm}$ and $t_{SiO_2}=\SI{0}{nm}$ and efficiency of $\varepsilon=0.69$ and use the first approach of modelling the efficiency presented in section \ref{Modelling the efficiency}.

%\begin{figure}[H]
	%\advance\leftskip-4cm
%	\begin{subfigure}{0.25\textwidth}
%		\centering
		%\includegraphics[width= 1 \textwidth]{GaAsnanoDBR2.eps}
%		\includegraphics[width= 1 \textwidth]{efficiency_initial_D246_t0_NA75.eps}
%		\caption{}
%		\label{efficiency_initial_D246_t0_NA75}
%	\end{subfigure}
%	\begin{subfigure}{0.25\textwidth}
%		\centering
		%\includegraphics[width= 1 \textwidth]{GaAsnanoDBR2.eps}
%		\includegraphics[width= 1 \textwidth]{efficiency_final_D246_t0_NA75.eps}
%		\caption{}
%		\label{efficiency_final_D246_t0_NA75}
%	\end{subfigure}
%	\caption{Structure A. (a) $\varepsilon$ ($NA=75$) as a function of the initial coefficients expressed with the propagation constant $\beta$. (b) $\varepsilon$ ($NA=75$) as a function of the final coefficients expressed with the propagation constant $\beta$.}
%	\label{efficiency_D246_t0_NA75}
%\end{figure}

\begin{figure}[ht]
	%\advance\leftskip-4cm
	\begin{subfigure}{0.5\textwidth}
		\centering
		\includegraphics[width= 1 \textwidth]{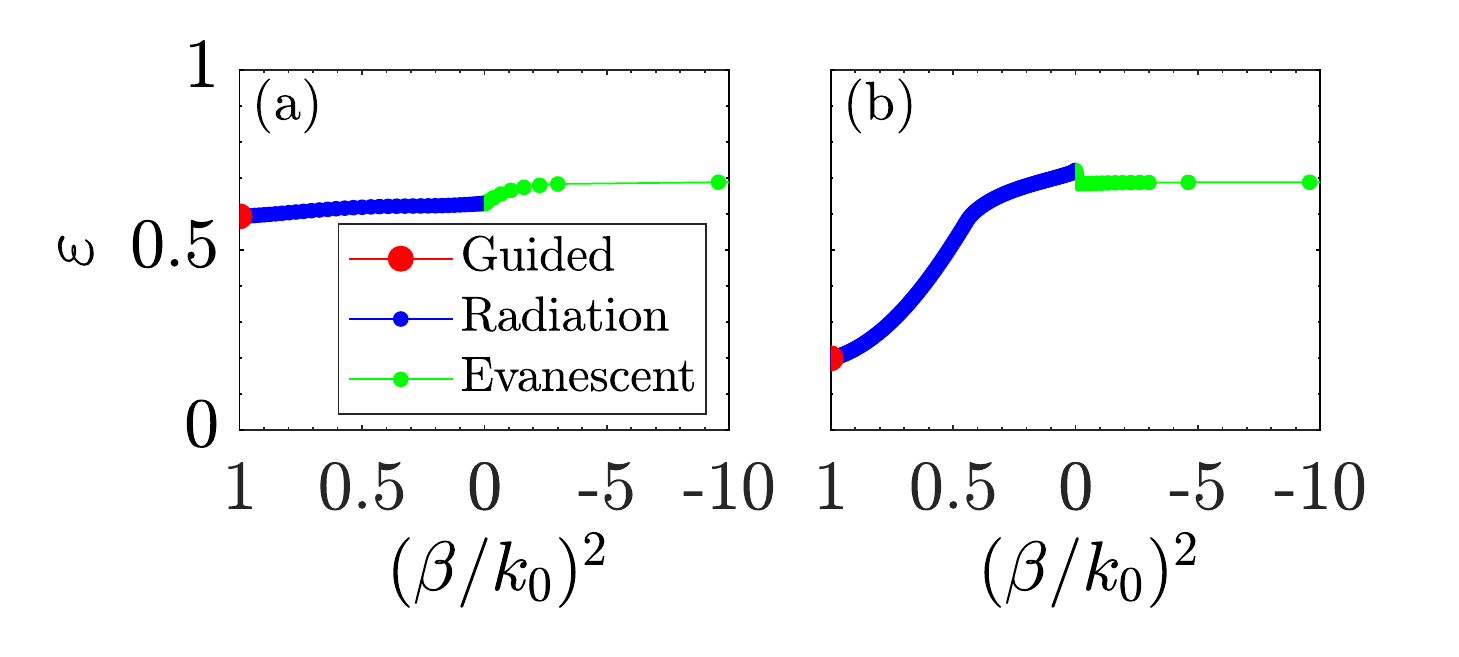}
	\end{subfigure}
	\caption{(a) Efficiency, $\varepsilon$ ($\mathrm{NA}=0.75$), as a function of the initial coefficients expressed with the propagation constant $(\beta/k_0)^2$. (b) Efficiency, $\varepsilon$ ($NA=75$), as a function of the final coefficients expressed with the propagation constant $(\beta/k_0)^2$.}
	\label{efficiency_D246_t0_NA75}
\end{figure}

We will now apply the modelling methods for the structure with the largest efficiency, $D=\SI{238}{nm}$ and $t_{\rm SiO_2}=\SI{0}{nm}$.
In Fig.\ (\ref{efficiency_D246_t0_NA75}) the efficiency is shown as a function of the initial coefficients (Fig.\ (\ref{efficiency_D246_t0_NA75}a)) and the final coefficients (Fig.\ (\ref{efficiency_D246_t0_NA75}b)), expressed with the propagation constant $(\beta/k_0)^2$. This corresponds to using the two different methods for calculating the efficiency presented in the previous subsection. The very first red point in Fig.\ (\ref{efficiency_D246_t0_NA75}a) corresponds to the fundamental mode (main channels) and includes all scattering channels of the fundamental mode ($c_1$, $c_2$ and $c_3$). This is sufficient to describe most of the efficiency. Then the initial background radiation modes (blue) are added one by one, starting from larger values of $(\beta/k_0)^2$, i.e. from predominantly vertical emission. This part of the curve is very flat, which means that the initial background radiation modes (channels originating from $\beta_t$ and $\beta_b$) are not crucial for the efficiency, which was also indicated by the low emission rates of the radiation modes in the infinite nanowire. Finally, the evanescent background modes (green) are included from small negative values of $(\beta/k_0)^2$, i.e. slowly decaying evanescent modes, to large negative values of $(\beta/k_0)^2$, i.e. fast decaying evanescent modes. Here there is a small increase due to the slowly decaying evanescent modes. Now in Fig.\ (\ref{efficiency_D246_t0_NA75}b), the very first red point also corresponds to the fundamental mode but only includes the direct transmission, $c_1$. Here the efficiency is only $\sim0.2$, much smaller than the total efficiency. Then the scattered radiation modes (blue) corresponding to $c_2$ and $c_3$ are added one by one, also starting from larger values of $(\beta/k_0)^2$, i.e. from predominantly vertical emission. Here there is a massive increase in the efficiency, which proves the importance of $c_2$ and $c_3$ to the efficiency. At some point, there is a kink in the blue part of the curve, which is due to the limited numerical aperture, as an $\rm NA$ of 0.75 corresponds to $(\beta/k_0)^2=0.36$. However, the remaining part of the curve is not completely flat, and this is due to the non-perfect transmission of the radiation modes of the nanowire to the radiation modes in the air. The radiation modes of the nanowire mainly transmit into radiation modes with the same value of $\beta$, but there is some scattering into the other radiation modes. Finally, there is a tiny decrease due to the evanescent modes, as very few of the evanescent modes transmit into radiation at the top interface. By comparing the evanescent parts of Fig.\ (\ref{efficiency_D246_t0_NA75}a) and Fig.\ (\ref{efficiency_D246_t0_NA75}b), we see that the effect of the evanescent modes is mainly back scattering into other modes at the interfaces rather than direct transmission scattering. To summarize, the key point in the comparison between Fig.\ (\ref{efficiency_D246_t0_NA75}a) and Fig.\ (\ref{efficiency_D246_t0_NA75}b) is that the scattering into radiation of the fundamental mode is crucial for the efficiency.

\begin{figure}[ht]
	%\advance\leftskip-4cm
	\begin{subfigure}{0.5\textwidth}
		\centering
		\includegraphics[width= 1 \textwidth]{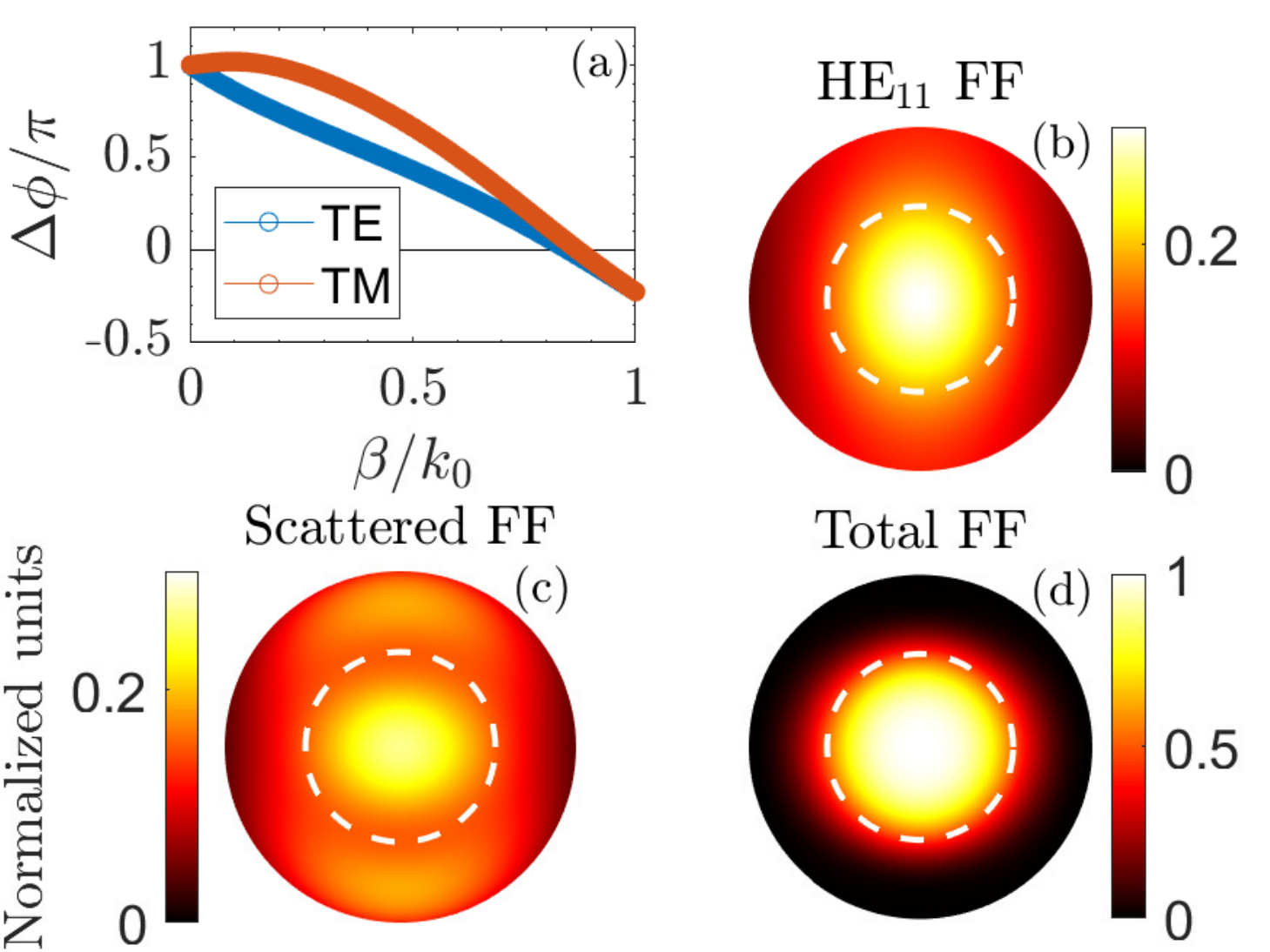}
	\end{subfigure}
	\caption{(a) The phase difference between the direct transmission of the fundamental mode and the background continuum for TE and TM modes as a function of the propagation constant, $\beta/k_0$. (b) The far-field of the fundamental mode. (c) The far-field of the background continuum. (d) The total far-field. The white dotted line indicates $\mathrm{NA}=0.75$. Be aware of the different color scales that have been used for the far-fields.}
	\label{farfield_D238_t0}
\end{figure}

%\begin{figure}[H]
	%\advance\leftskip-4cm
%	\begin{subfigure}{0.25\textwidth}
%		\centering
		%\includegraphics[width= 1 \textwidth]{GaAsnanoDBR2.eps}
%		\includegraphics[width= 1 \textwidth]{phase_D246_t0v2.eps}
%		\caption{}
%		\label{phase_D246_t0}
%	\end{subfigure}
%	\begin{subfigure}{0.25\textwidth}
%		\centering
		%\includegraphics[width= 1 \textwidth]{GaAsnanoDBR2.eps}
%		\includegraphics[width= 1 \textwidth]{farfield_HE11_D246_t0v2.eps}
%		\caption{}
%		\label{farfield_HE11_D246_t0}
%	\end{subfigure}
%	\begin{subfigure}{0.25\textwidth}
%		\centering
		%\includegraphics[width= 1 \textwidth]{GaAsnanoDBR2.eps}
%		\includegraphics[width= 1 \textwidth]{farfield_radiation_D246_t0v2.eps}
%		\caption{}
%		\label{farfield_radiation_D246_t0}
%	\end{subfigure}
%	\begin{subfigure}{0.25\textwidth}
%		\centering
		%\includegraphics[width= 1 \textwidth]{GaAsnanoDBR2.eps}
%		\includegraphics[width= 1 \textwidth]{farfield_total_D246_t0v2.eps}
%		\caption{}
%		\label{farfield_total_D246_t0}
%	\end{subfigure}
%	\caption{Structure A. (a) The phase difference between the direct transmission of the fundamental mode and the background continuum for TE and TM modes as a function of the propagation constant. (b) The far-field of the fundamental mode. (c) The far-field of the background continuum. (d) The total far-field. Be aware of the different color scales that have been used for the far-fields.}
%	\label{farfield_D246_t0}
%\end{figure}

Now we will visualize the interference between the direct transmission of the fundamental mode and the scattered channels by inspecting the phase changes and the far-fields. 
In Fig.\ (\ref{farfield_D238_t0}a), the phase difference in the air layer between the direct transmission of the fundamental mode and the entire background is shown as a function of the propagation constant for TE and TM modes. For the light that propagates vertically, the phase difference is close to zero, such that there is constructive interference between the direct transmission and the background radiation. For the light that propagates horizontally, the phase difference is closer to $\pi$, and thus there is destructive interference. In Fig.\ (\ref{farfield_D238_t0}b), Fig.\ (\ref{farfield_D238_t0}c) and Fig.\ (\ref{farfield_D238_t0}d) the far-fields of the direct transmission of the fundamental mode, the background radiation and the total field are shown. Here we can directly observe the effect caused by the phase difference. In the center of the total far-field, the field is enhanced due to the constructive interference, but for the light that propagates horizontally there is destructive interference. As such, the interference between the direct emission and the radiation focuses the far-field.% Interference between the direct transmission and the scattered background has been documented before, where ring-like patterns could be observed in the far-field \cite{AndreasPHD}.  

A similar analysis of the efficiency for the structure with the largest Purcell factor is included in Supplementary \ref{sec:eff max Fp}.% Here the far-field of the background is much 

\subsection{Enhanced Purcell factor}

As shown in Fig.\ (\ref{Fp_result}), there are deviations between the full model and the SMM for the Purcell factor, and we wish to understand where these deviations appear. Therefore we will introduce a model which can identify where these deviations appear and apply the model on the structure with the largest Purcell factor, namely $D=\SI{250}{nm}$ and $t_{\rm SiO_2}=\SI{13}{nm}$.

\subsubsection{Purcell factor contributions from the emission channels}

To gain physical insight into the physics of the Purcell factor, we will use a model which stepwise increases the complexity. At each step the Purcell factor is calculated $F_{\rm p}=P_{\rm T}/P_{\rm Bulk}$ along with the power into the fundamental mode, $P_{\mathrm{HE}_{11}}/P_{\rm Bulk}$. The starting point is the SMM, where only the fundamental mode is included; then, in seven steps, the complexity increases until the full model is reached. Specifically, the initial inputs ($\mathbf{a}_{\infty \rm NW}$ and $\mathbf{b}_{\infty \rm NW}$) and the top and bottom reflections ($\mathbf{R}_{\rm top}$ and $\mathbf{R}_{\rm bot}$) will be manipulated at each step. Each step has a direct physical interpretation. We will now list the 7 steps in the model and for each step, write up the physical effect that is now included. This means that for each step in the model, all previous effects are also included.\begin{enumerate}[noitemsep]
    \item SMM
    \item Scattering of the fundamental mode at the top interface
    \item Scattering of the fundamental mode at the bottom interface
    \item Back-scattering of the background continuum to the fundamental mode at the top interface
    \item Back-scattering of the background continuum to the fundamental mode at the bottom interface
    \item Scattering of the background continuum to itself at both interfaces
    \item Including initial background continuum
\end{enumerate} Steps number 4 and 5 correspond to the process $\textrm{HE}_{11}\rightarrow\textrm{radiation/evanescent}\rightarrow\textrm{HE}_{11}$ which is a recycling effect, and we will show the importance of this process for the Purcell factor. Step number 6 corresponds to the process $\textrm{radiation/evanescent}\rightarrow\textrm{radiation/evanescent}$, which also opens up for further scattering channels such as $\textrm{radiation/evanescent}\rightarrow\textrm{radiation/evanescent}\rightarrow\textrm{HE}_{11}$.

%$\textrm{HE}_{11}\rightarrow\textrm{radiation/evanescent}\rightarrow\textrm{HE}_{11}$. This process is a recycling effect which is important for the Purcell factor and later on we will present models which can quantify the contributions of the different channels and mode coupling, including this recycling effect.
%1. 1D Fabry-Pérot model\newline
%2. Scattering of the fundamental mode at the top interface\newline
%3. Scattering of the fundamental mode at the bottom interface\newline
%4. Back-scattering of the background continuum to the fundamental mode at the top interface\newline
%5. Back-scattering of the background continuum to the fundamental mode at the bottom interface\newline
%6. Scattering of the background continuum to itself at both interfaces\newline
%7. Including initial background continuum\newline
The 7 steps can then be translated to the initial inputs and the top and bottom reflections in the following schematic way:

%\begin{equation}
%	\begin{split}
%		\mathbf{a}_{\infty \rm NW}&=
%		\begin{bmatrix}
%			1 & 7 & \cdots & 7
%		\end{bmatrix}\\
%		\mathbf{b}_{\infty \rm NW}&=
%		\begin{bmatrix}
%			1 & 7 & \cdots & 7
%		\end{bmatrix}\\
%		\mathbf{R}_{\rm top}&=
%		\begin{bmatrix}
%			1 & 4 & 4 & \dots  &4 \\
%			2 & 2 & 6 & \dots  & 6 \\
%			2 & 6 & 2 & \dots  & 6 \\
%			\vdots & \vdots & \vdots & \ddots & \vdots \\
%			2 & 6 & 6 & \dots  & 2
%		\end{bmatrix}_{\rm top}\\
%		\mathbf{R}_{\rm bot}&=
%		\begin{bmatrix}
%			1 & 5 & 5 & \dots  &5 \\
%			3 & 2 & 6 & \dots  & 6 \\
%			3 & 6 & 2 & \dots  & 6 \\
%			\vdots & \vdots & \vdots & \ddots & \vdots \\
%			3 & 6 & 6 & \dots  & 2
%		\end{bmatrix}_{\rm bot}
%	\end{split}
%\end{equation}

\begin{equation}
	\begin{split}
		\mathbf{a}_{\infty \rm NW}^{(\alpha)}&=
		\begin{bmatrix}
			1 & 7 & \cdots & 7
		\end{bmatrix}\\
	\end{split}
\end{equation}
\begin{equation}
	\begin{split}
		\mathbf{b}_{\infty \rm NW}^{(\alpha)}&=
		\begin{bmatrix}
			1 & 7 & \cdots & 7
		\end{bmatrix}\\
	\end{split}
\end{equation}
\begin{equation}
	\begin{split}
		\mathbf{R}_{\rm top}^{(\alpha)}&=
		\begin{bmatrix}
			1 & 4 & 4 & \dots  &4 \\
			2 & 2 & 6 & \dots  & 6 \\
			2 & 6 & 2 & \dots  & 6 \\
			\vdots & \vdots & \vdots & \ddots & \vdots \\
			2 & 6 & 6 & \dots  & 2
		\end{bmatrix}_{\rm top}\\
	\end{split}
\end{equation}
\begin{equation}
	\begin{split}
		\mathbf{R}_{\rm bot}^{(\alpha)}&=
		\begin{bmatrix}
			1 & 5 & 5 & \dots  &5 \\
			3 & 2 & 6 & \dots  & 6 \\
			3 & 6 & 2 & \dots  & 6 \\
			\vdots & \vdots & \vdots & \ddots & \vdots \\
			3 & 6 & 6 & \dots  & 2
		\end{bmatrix}_{\rm bot}
	\end{split}
\end{equation}

Here, the superscript $\alpha$ represents the seven complexity steps.
For a given step $\alpha$, each matrix entry $> \alpha$ is set to zero, while entries $\leq \alpha$ keep their original value. We will use the step  $\alpha=4$ as an example:

%\begin{equation}
%	\begin{split}
%		\mathbf{a}_{\infty \rm NW}&=
%		\begin{bmatrix}
%			a_{1} & 0 & \cdots & 0
%		\end{bmatrix}\\
%		\mathbf{b}_{\infty \rm NW}&=
%		\begin{bmatrix}
%			b_{1} & 0 & \cdots & 0
%		\end{bmatrix}\\
%		\mathbf{R}_{\rm top}&=
%		\begin{bmatrix}
%			r_{11} & r_{12} & r_{13} & \dots  &r_{1N} \\
%			r_{21} & r_{22} & 0 & \dots  & 0 \\
%			r_{31} & 0 & r_{33} & \dots  & 0 \\
%			\vdots & \vdots & \vdots & \ddots & \vdots \\
%			r_{N1} & 0 & 0 & \dots  & r_{NN}
%		\end{bmatrix}_{\rm top}\\
%		\mathbf{R}_{\rm bot}&=
%		\begin{bmatrix}
%			r_{11} & 0 & 0 & \dots  &0 \\
%			r_{21} & r_{22} & 0 & \dots  & 0 \\
%			r_{31} & 0 & r_{33} & \dots  & 0 \\
%			\vdots & \vdots & \vdots & \ddots & \vdots \\
%			r_{N1} & 0 & 0 & \dots  & r_{NN}
%		\end{bmatrix}_{\rm bot}
%	\end{split}
%\end{equation}

\begin{equation}
	\begin{split}
		\mathbf{a}_{\infty \rm NW}^{(4)}&=
		\begin{bmatrix}
			a_{1} & 0 & \cdots & 0
		\end{bmatrix}
	\end{split}
\end{equation}
\begin{equation}
	\begin{split}
		\mathbf{b}_{\infty \rm NW}^{(4)}&=
		\begin{bmatrix}
			b_{1} & 0 & \cdots & 0
		\end{bmatrix}
	\end{split}
\end{equation}
\begin{equation}
	\begin{split}
		\mathbf{R}_{\rm top}^{(4)}&=
		\begin{bmatrix}
			r_{11} & r_{12} & r_{13} & \dots  &r_{1N} \\
			r_{21} & r_{22} & 0 & \dots  & 0 \\
			r_{31} & 0 & r_{33} & \dots  & 0 \\
			\vdots & \vdots & \vdots & \ddots & \vdots \\
			r_{N1} & 0 & 0 & \dots  & r_{NN}
		\end{bmatrix}_{\rm top}\\
	\end{split}
\end{equation}
\begin{equation}
	\begin{split}
		\mathbf{R}_{\rm bot}^{(4)}&=
		\begin{bmatrix}
			r_{11} & 0 & 0 & \dots  &0 \\
			r_{21} & r_{22} & 0 & \dots  & 0 \\
			r_{31} & 0 & r_{33} & \dots  & 0 \\
			\vdots & \vdots & \vdots & \ddots & \vdots \\
			r_{N1} & 0 & 0 & \dots  & r_{NN}
		\end{bmatrix}_{\rm bot}
	\end{split}
\end{equation}

At each step in the previous model building, the entire background continuum was added, i.e. the entire column or row was added at once. As such, it is difficult to quantify which part of the background continuum that is important. Therefore we will also present the models where the elements for the background continuum are increased one at a time. In this way it can be quantified how the different parts of the background continuum contribute to the Purcell factor.

\subsubsection{Influence of radiation modes on the Purcell factor}

%\paragraph{Purcell factor.}

\begin{figure}[b]
	%\advance\leftskip-4cm
	\begin{subfigure}{0.5\textwidth}
		\centering
		\includegraphics[width= 1 \textwidth]{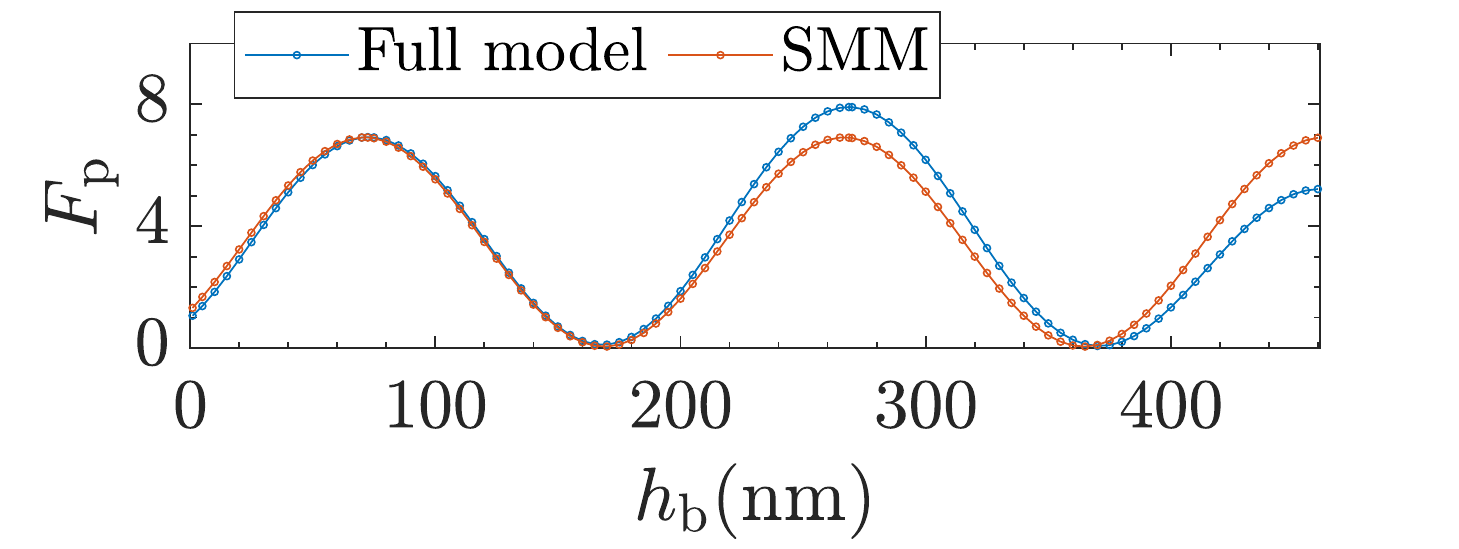}
	\end{subfigure}
	\caption{Purcell factor, $F_{\rm p}$, computed using the full model and SMM as a function of the dipole position from the bottom interface, $h_{\rm b}$. $D=\SI{250}{nm}$ and $t_{\rm SiO_2}=\SI{13}{nm}$.}
	\label{Fp_hb_D246_t12}
\end{figure}

We will now apply the model for the Purcell factor for the structure with the largest Purcell factor, $D=\SI{250}{nm}$ and $t_{\rm SiO_2}=\SI{13}{nm}$.
In Fig.\ (\ref{Fp_hb_D246_t12}), the Purcell factor is shown as a function of the dipole position from the bottom, $h_{\rm b}$. The agreement between the SMM and the full model for the 1st antinode is good, but there are deviations for the 2nd and 3rd antinodes. Nevertheless, we will focus on the analysis of the 1st and the 2nd antinode.

% \begin{figure}[H]
% 	%\advance\leftskip-4cm
% 	\begin{subfigure}{0.25\textwidth}
% 		\centering
% 		%\includegraphics[width= 1 \textwidth]{GaAsnanoDBR2.eps}
% 		\includegraphics[width= 1 \textwidth]{Fp_models_1st_D246_t12.eps}
% 		\caption{}
% 		\label{Fp_models_1st_D246_t12}
% 	\end{subfigure}
% 	\begin{subfigure}{0.25\textwidth}
% 		\centering
% 		%\includegraphics[width= 1 \textwidth]{GaAsnanoDBR2.eps}
% 		\includegraphics[width= 1 \textwidth]{Fp_models_2nd_D246_t12.eps}
% 		\caption{}
% 		\label{Fp_models_2nd_D246_t12}
% 	\end{subfigure}
% 	%\caption{.}
% 	%\label{Fp_models_D246_t12}
% 	\begin{subfigure}{0.25\textwidth}
% 		\centering
% 		%\includegraphics[width= 1 \textwidth]{GaAsnanoDBR2.eps}
% 		\includegraphics[width= 1 \textwidth]{Fp_models_1st_D246_t12_continuum.eps}
% 		\caption{}
% 		\label{Fp_models_1st_D246_t12_c}
% 	\end{subfigure}
% 	\begin{subfigure}{0.25\textwidth}
% 		\centering
% 		%\includegraphics[width= 1 \textwidth]{GaAsnanoDBR2.eps}
% 		\includegraphics[width= 1 \textwidth]{Fp_models_2nd_D246_t12_continuum.eps}
% 		\caption{}
% 		\label{Fp_models_2nd_D246_t12_c}
% 	\end{subfigure}
% 	\caption{Purcell factor ($P_{\rm T}/P_{\rm Bulk}$) and fundamental mode enhancement ($P_{\mathrm{HE}_{11}}/P_{\rm Bulk}$) for the 1st (a) and 2nd (b) antinode as a function of the model number. In (c) and (d) the background continuum is continuously included between each model number.}
% 	\label{Fp_models_D246_t12}
% \end{figure}

\begin{figure}[ht]
	%\advance\leftskip-4cm
	\begin{subfigure}{0.5\textwidth}
		\centering
		\includegraphics[width= 1 \textwidth]{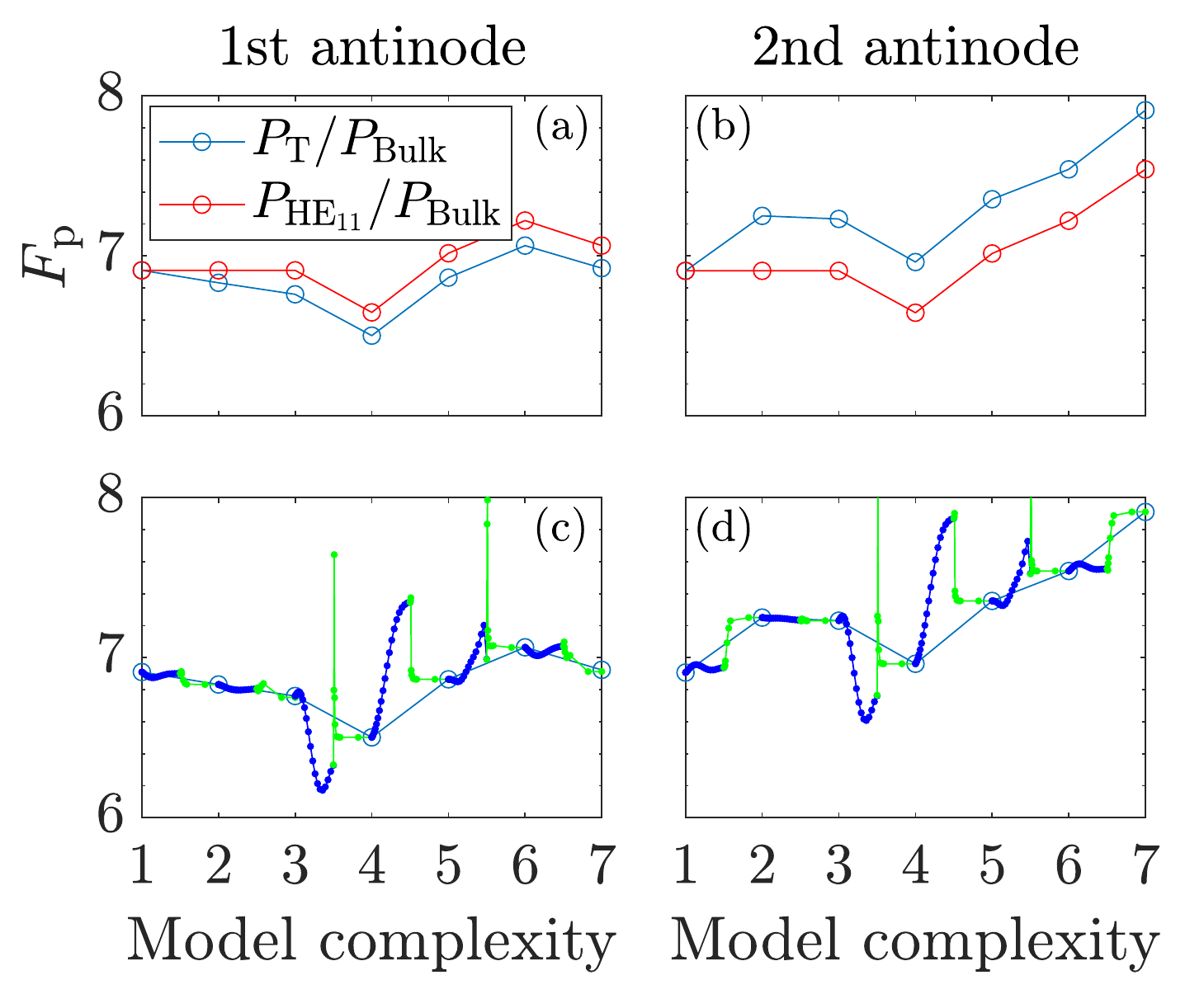}
	\end{subfigure}
	\caption{Purcell factor ($P_{\rm T}/P_{\rm Bulk}$) and fundamental mode enhancement ($P_{\mathrm{HE}_{11}}/P_{\rm Bulk}$) for the 1st (a) and 2nd (b) antinode as a function of the model complexity progression. In (c) and (d) the background continuum is continuously included between each model complexity.}
	\label{Fp_models_D250_t13}
\end{figure}

In Fig.\ (\ref{Fp_models_D250_t13}a) and Fig.\ (\ref{Fp_models_D250_t13}b), the Purcell factor and the power enhancement of the fundamental mode are shown as a function of the model complexity progression for the 1st and 2nd antinode (model complexity number $\alpha$ will be shortened n. $\alpha$). Evidently, the analysis of the Purcell factor is complicated due to contributions of the entire background continuum, multiple scattering channels and feedback mechanism. Therefore there are changes in the Purcell factor for all steps in the model complexity, which makes it challenging to model the Purcell factor using only a few modes. The continuum of radiation modes can be modelled using leaky modes, which can enable the modelling using only a few modes. This has been demonstrated in photonic crystal microcavities where strong feedback mechanisms also were present \cite{Lalanne2004}. However, by using the presented model, we will obtain an in-depth physical insight.

The Purcell factor starts at the same value with the SMM for both antinodes. By including scattering of the fundamental mode at the top interface (n.\ 2) there is a significant increase for the 2nd antinode but a very small decrease for the 1st antinode. Already now, the deviations compared to the SMM have started to appear. By including the scattering at the bottom interface (n.\ 3) the 2nd antinode is almost unaffected, but a small decrease appears for the 1st antinode. Now including the back-scattering at the top interface (n.\ 4), there is a large decrease in the Purcell factor at both antinodes. This decrease is directly represented in $P_{\mathrm{HE}_{11}}/P_{\rm Bulk}$. Interestingly, when including the back-scattering at the bottom interface (n.\ 5), there is now a large increase in the Purcell factor at both antinodes, which is also represented in $P_{\mathrm{HE}_{11}}/P_{\rm Bulk}$. This also shows that the recycling effect $\textrm{HE}_{11}\rightarrow\textrm{radiation/evanescent}\rightarrow\textrm{HE}_{11}$ can provide both negative and positive contributions. When the background is allowed to scatter to itself, i.e. $\textrm{radiation/evanescent}\rightarrow\textrm{radiation/evanescent}$ (n.\ 6), there is an increase for both antinodes. This increase is directly represented in $P_{\mathrm{HE}_{11}}/P_{\rm Bulk}$, which in fact means that the process $\textrm{radiation/evanescent}\rightarrow\textrm{radiation/evanescent}\rightarrow\textrm{HE}_{11}$ is dominating compared to $\textrm{radiation/evanescent}\rightarrow\textrm{radiation/evanescent}$ itself. Finally, by including the initial background (n.\ 7), there is an increase for the 2nd antinode but a decrease for the 1st antinode. These changes also correspond to the change in $P_{\mathrm{HE}_{11}}/P_{\rm Bulk}$, which means it is the process of $\textrm{radiation/evanescent}\rightarrow\textrm{HE}_{11}$ that is important for the initial background.

The main differences between the two antinodes appear, when the fundamental mode scatters at the top interface (n.\ 2) and when the initial background (n.\ 7) is included. To better understand which part of the background continuum is important, we will also consider the continuous steps of the model complexity. This is now shown in Fig.\ (\ref{Fp_models_D250_t13}c) and Fig.\ (\ref{Fp_models_D250_t13}d). Here we observe that the main positive contributions at the 2nd antinode are due to the slowly decaying evanescent modes. This is the process of $\textrm{HE}_{11}\rightarrow\textrm{evanescent}$ at the top interface and the process of the initial background $\textrm{evanescent}\rightarrow\textrm{HE}_{11}$. For the back-scattering at the interfaces (n.\ 4 and n.\ 5), we observe that the propagating radiation modes can also significantly affect the Purcell factor. 

An additional important observation is that $P_{\mathrm{HE}_{11}}/P_{\rm Bulk}$ exceeds $P_{\rm T}/P_{\rm Bulk}$ for the first antinode due to the negative contributions. This would in fact result in $\beta$ factors above 1 using the definition $\beta_{\mathrm{HE}_{11}}=P_{\mathrm{HE}_{11}}/P_{\rm T}$. This indicates that the $\beta$ factor might not be a suitable figure of merit for structures where the SMM breaks down, or at least one should be very careful in the definition of the $\beta$ factor. Alternatively, the typical interpretation of the $\beta$ factor as a standard power fraction should be reconsidered in the regime of the SMM breakdown.

The analysis of the Purcell factor for the structure with the largest efficiency, $D=\SI{238}{nm}$ and $t_{\rm SiO_2}=\SI{0}{nm}$, is included in Supplementary \ref{sec:Fp max eff}. %Here we show that the back-scattering at both interfaces (n. 4 and n. 5) provide a negative contribution. This is caused by the change in the silica layer thickness.  

\subsection{Wavelength dependence}

In this section, we will present the broadband performance of the nanopost. The focus will be on the designs with the largest Purcell factor and efficiency, respectively.   

%In the previous work on the nanowire optical nanocavity, the broadband performance was already demonstrated for the structure with $D=\SI{246}{nm}$ and $t_{SiO_2}=\SI{8}{nm}$ at a resonance of $\lambda_c=\SI{928}{nm}$ \cite{oldnanopost}. 

%\begin{figure}[H]
	%\advance\leftskip-4cm
%	\begin{subfigure}{0.25\textwidth}
%		\centering
		%\includegraphics[width= 1 \textwidth]{GaAsnanoDBR2.eps}
%		\includegraphics[width= 1 \textwidth]{Fp_D250_t13_lambda.eps}
%		\caption{}
%		\label{Fp_D250_t13_lambda}
%	\end{subfigure}
%	\begin{subfigure}{0.25\textwidth}
%		\centering
		%\includegraphics[width= 1 \textwidth]{GaAsnanoDBR2.eps}
%		\includegraphics[width= 1 \textwidth]{Fp_D238_t0_lambda.eps}
%		\caption{}
%		\label{Fp_D238_t0_lambda}
%	\end{subfigure}
%	\begin{subfigure}{0.25\textwidth}
%		\centering
		%\includegraphics[width= 1 \textwidth]{GaAsnanoDBR2.eps}
%		\includegraphics[width= 1 \textwidth]{eff_D250_t13_lambda.eps}
%		\caption{}
%		\label{eff_D250_t13_lambda}
%	\end{subfigure}
%	\begin{subfigure}{0.25\textwidth}
%		\centering
		%\includegraphics[width= 1 \textwidth]{GaAsnanoDBR2.eps}
%		\includegraphics[width= 1 \textwidth]{eff_D238_t0_lambda.eps}
%		\caption{}
%		\label{eff_D238_t0_lambda}
%	\end{subfigure}
%	\caption{(a) and (b) Purcell factor for the two antinodes for the two structures as a function of wavelength. (c) and (d) efficiency for the two antinodes for the two structures as a function of wavelength.}
%\end{figure}

\begin{figure}[tb]
	%\advance\leftskip-4cm
	\begin{subfigure}{1\linewidth}
		\centering
		\includegraphics[width= 1 \linewidth]{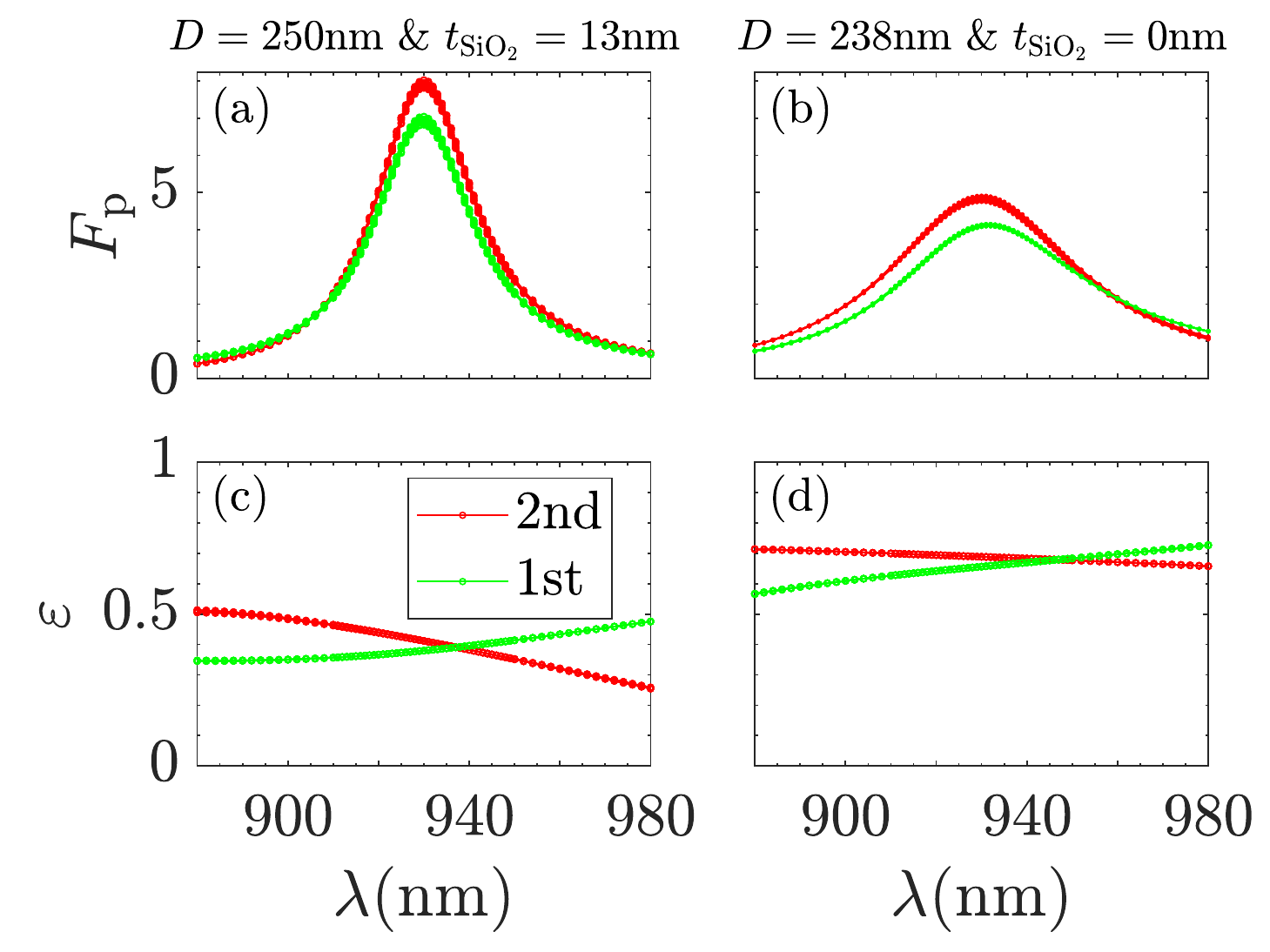}
	\end{subfigure}
	\caption{(a) and (b) Purcell factor, $F_{\rm p}$, of the two antinodes for the two structures as a function of wavelength, $\lambda$. (c) and (d) efficiency, $\varepsilon$ ($\mathrm{NA}=0.75$), of the two antinodes for the two structures as a function of wavelength, $\lambda$. The numerical uncertainty is represented by the thickness of the curves.}
	\label{Fp_eff_lambda}
\end{figure}

In Fig.\ (\ref{Fp_eff_lambda}a), the Purcell factor is shown as a function of the wavelength for the 1st and 2nd antinode of the structure with the largest Purcell factor. The Purcell factor of the 2nd antinode performs better than the 1st antinode close to the resonance wavelength, and the spectral width at FWHM (full width half maximum) of the antinodes are approximately $\Delta\lambda_{\mathrm{2nd}}=\SI{27}{nm}$ and $\Delta\lambda_{\mathrm{1st}}=\SI{28}{nm}$ showcasing the broadband performance. The spectrum for the two antinodes is not completely symmetric, and the two curves for the antinodes also cross further away from the resonance. In Fig.\ (\ref{Fp_eff_lambda}b), the Purcell factor is shown as a function of the wavelength for the 1st and 2nd antinode of the structure with the largest efficiency (at resonance). Here the Purcell factor is much lower and the spectral width much broader at $\Delta\lambda_{\mathrm{2nd}}=\SI{52}{nm}$ and $\Delta\lambda_{\mathrm{1st}}=\SI{56}{nm}$. Furthermore, the resonance wavelength for the 1st antinode is slightly shifted to $\lambda_{\rm r}=\SI{931.5}{nm}$. This can be explained by the low Q factor and the neighbouring low Q factor cavity modes, i.e. the cavity modes with 2 and 4 antinodes. Due to the low Q factor, there is a small spectral overlap causing the slight shift. In the Supplementary \ref{sec:asym}, we present a nanopost design where the resonance wavelength shift is more pronounced between the two antinodes.  

In Fig.\ (\ref{Fp_eff_lambda}c) and Fig.\ (\ref{Fp_eff_lambda}d) the efficiencies ($\mathrm{NA}=0.75$) are shown as a function of the wavelength for the 1st and 2nd antinode for the two structures. The main characteristic of the efficiency is that it does not follow the Purcell factor, which is surprising compared to traditional Fabry-Pérot cavities. Instead, the efficiency changes roughly linearly across the resonance, and in general, the slope for the 2nd antinode is negative and positive for the 1st antinode. This means that the maximum efficiency is in fact achieved off-resonance, $\varepsilon_{\rm 2nd, NA=0.75}(\lambda=\SI{880}{nm})=0.71$, but at a much smaller Purcell factor.

%Again, this shows that the efficiency can be increased even further by optimizing the design towards larger efficiency, but a the cost of a reduced Purcell factor.    

% Further investigation needs to be done to understand this behavior. 

\subsubsection{Analysis of the broadband collection efficiency}

In the previous work on the nanopost\cite{Kotal2021}, the broadband efficiency was attributed to the broadband $\beta$ factor, which again was attributed to the dielectric screening effect. This can also be seen in Fig.\ (\ref{power_nanowire}), where the emission into radiation modes is suppressed in most of the interval. Here we will define the $\beta$ factor for the fundamental mode as $\beta_{\mathrm{HE}_{11}}=P_{\mathrm{HE}_{11}}/P_{\rm T}$, even though it exceeds 1 as we have already shown. The focus is on the structure with the largest efficiency at resonance, $D=\SI{238}{nm}$ and $t_{\rm SiO_2}=\SI{0}{nm}$. 

\begin{figure}[tb]
	%\advance\leftskip-4cm
	\begin{subfigure}{1\linewidth}
		\centering
		\includegraphics[width= 1 \linewidth]{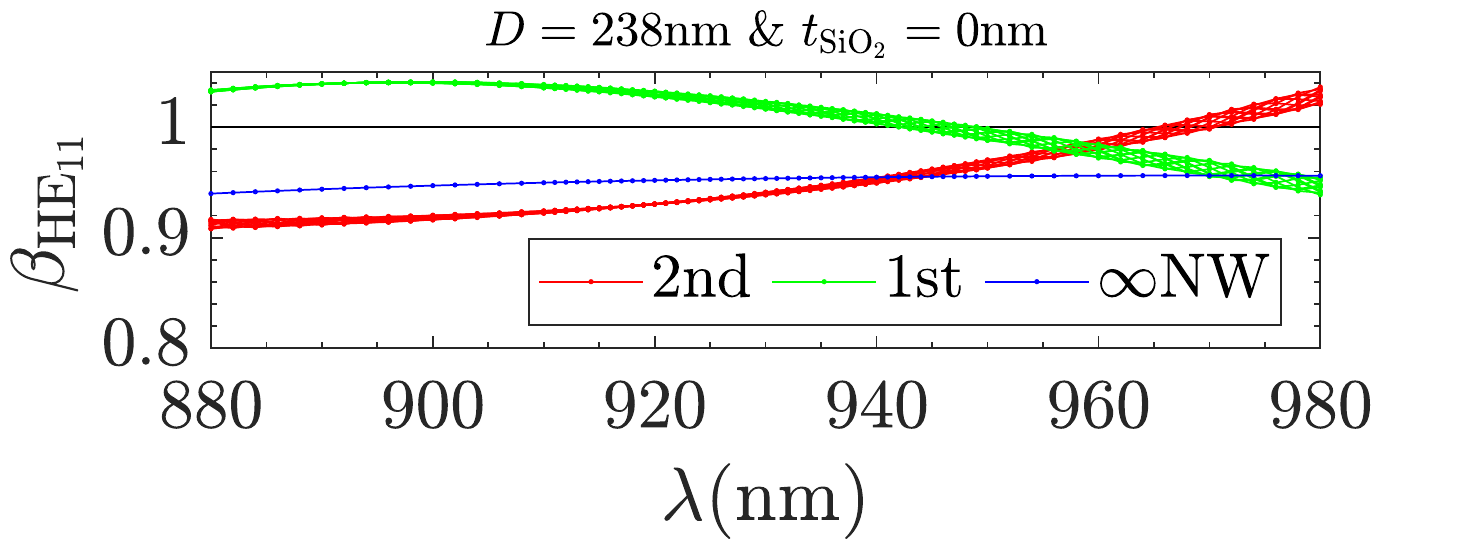}
	\end{subfigure}
	\caption{$\beta_{\mathrm{HE}_{11}}=P_{\mathrm{HE}_{11}}/P_{\rm T}$ as a function of the wavelength, $\lambda$, for the 2nd and 1st antinode and the infinite nanowire. The numerical uncertainty is represented by the thickness of the curves.}
	\label{betaHE11}
\end{figure}

In Fig.\ (\ref{betaHE11}), $\beta_{\mathrm{HE}_{11}}$ is shown as a function of the wavelength for the 2nd and 1st antinode and the infinite nanowire. $\beta_{\mathrm{HE}_{11}}$ is close to 1 (both above and below) in the entire interval, which indicates that the fundamental mode is still the dominating contribution. Due to the broadband $\beta_{\mathrm{HE}_{11}}$ of the infinite nanowire, it is not required to be on resonance to obtain large values of $\beta_{\mathrm{HE}_{11}}$, and the QD would need to be very close to a node before $\beta_{\mathrm{HE}_{11}}$ would decrease. Regardless, the fundamental mode scatters into radiation which affects $P_{\rm T}$ such that $\beta_{\mathrm{HE}_{11}}$ does not follow a Lorentzian curve like the Purcell factor.   

%Exactly due to the dielectric screening effect and thus large $\beta_{\mathrm{HE}_{11}}$ of the infinite nanowire, it is not required to be on resonance for the fundamental mode to be dominating. The QD needs to be very close to a node before the fundamental mode contribution decreases. Anyway, the fundamental mode scatters into radiation which affects $P_{\rm T}$ such that $\beta_{\mathrm{HE}_{11}}$ does not follow a Lorentzian curve like the Purcell factor.   

\begin{figure}[tb]
	%\advance\leftskip-4cm
	\begin{subfigure}{1\linewidth}
		\centering
		\includegraphics[width= 1 \linewidth]{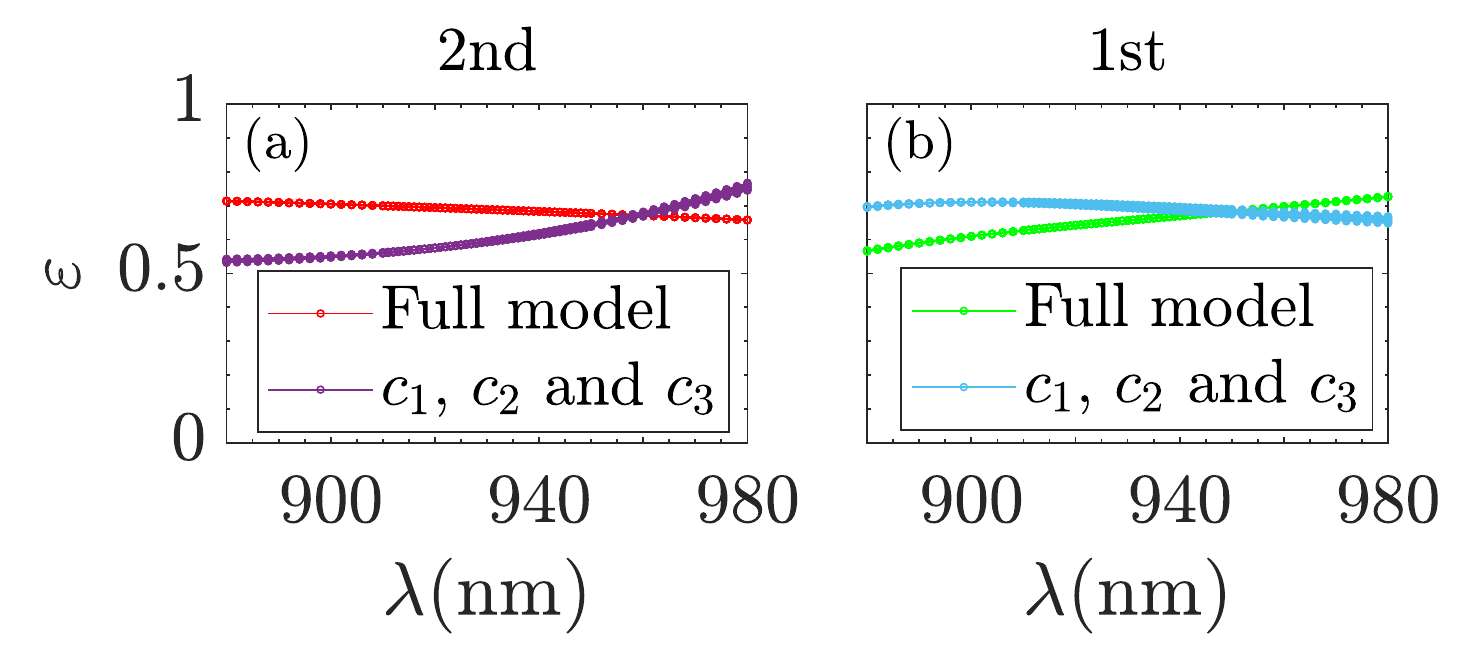}
	\end{subfigure}
	\caption{Efficiency, $\varepsilon$ ($\mathrm{NA}=0.75$), as a function of the wavelength, $\lambda$, for the 2nd antinode (a) and 1st antinode (b). The full model is compared to only using $c_1$, $c_2$ and $c_3$. $D=\SI{238}{nm}$ and $t_{\rm SiO_2}=\SI{0}{nm}$. The width of the curves represents the numerical uncertainty.}
	\label{eff_lambda_D238_models}
\end{figure}

To quantify how dominating the fundamental mode is for the efficiency, we will use the first method presented in Section \ref{Modelling the efficiency} to separate the main channels ($c_1$, $c_2$ and $c_3$) from the background channels.    
In Fig.\ (\ref{eff_lambda_D238_models}), we compare the efficiency of the full model to only including $c_1$, $c_2$ and $c_3$ for both antinodes. By comparing to Fig.\ (\ref{betaHE11}) we observe that the curves for the main channels directly follow the trend of $\beta_{\mathrm{HE}_{11}}$. However, there is a discrepancy between the efficiency of the full model and only using the main channels of $c_1$, $c_2$ and $c_3$. This discrepancy is not at a minimum on resonance ($\lambda=\SI{930}{nm}$), but almost at the minimum when $\beta_{\mathrm{HE}_{11}}=1$ for both antinodes. We observe that for $\beta_{\mathrm{HE}_{11}}<1$, the background channels provide a positive contribution to the efficiency, while for $\beta_{\mathrm{HE}_{11}}>1$ the background channels provide a negative contribution. This shows that the $\beta$ factor is still useful in the analysis but not necessarily a figure of merit for SPSs in the breakdown of the SMM. Furthermore, Fig.\ (\ref{eff_lambda_D238_models}) shows that the background channels still interfere with the main channels causing this discrepancy and even changing the slope of the curves for the efficiency.
As an example we will consider two wavelengths for the 2nd antinode and use the first method presented in Section \ref{Modelling the efficiency}:  

%This discrepancy is almost at a minimum when $\beta_{\mathrm{HE}_{11}}=1$ for both antinodes and we observe that for $\beta_{\mathrm{HE}_{11}}<1$ the background channels provide a positive contribution to the efficiency, while for $\beta_{\mathrm{HE}_{11}}>1$ the background channels provide a negative contribution. This shows that the background channels still interfere with the main channels causing this discrepancy and even changes the slope of the curves for the efficiency.
%As an example we will consider two wavelengths for the 2nd antinode and use the first model presented in Section \ref{Modelling the efficiency}. 

%This discrepancy is not exactly at a minimum on resonance ($\lambda=\SI{930}{nm}$) nor when $\beta_{\mathrm{HE}_{11}}=1$. However, as a trend the discrepancy increases far away from the resonance. This shows that the background channels still interfere with the main channels causing this discrepancy and even changes the slope of the curves for the efficiency.
%As an example we will consider two wavelengths for the 2nd antinode and use the first method of the first approach presented in Section \ref{Modelling the efficiency}.

\begin{figure}[tb]
	%\advance\leftskip-4cm
	\begin{subfigure}{1\linewidth}
		\centering
		\includegraphics[width= 1 \linewidth]{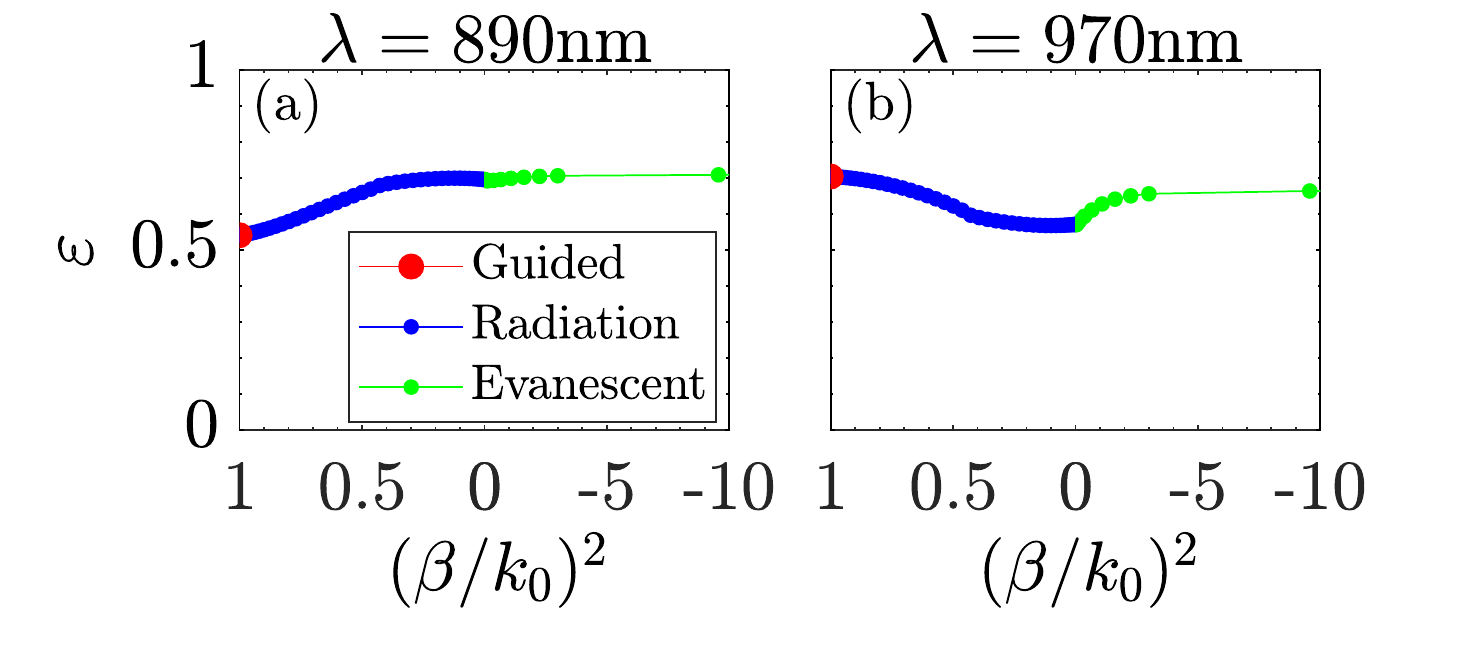}
	\end{subfigure}
	\caption{Efficiency, $\varepsilon$ ($\mathrm{NA}=0.75$), as a function of the initial coefficients expressed with the propagation constant $(\beta/k_0)^2$ for $\lambda=\SI{890}{nm}$ (a) and $\lambda=\SI{970}{nm}$ (b) at the 2nd antinode.}
	\label{eff_D238_t0_lambda890_970}
\end{figure}

In Fig.\ (\ref{eff_D238_t0_lambda890_970}) the efficiency is shown as a function of the initial coefficients expressed with the propagation constant $(\beta/k_0)^2$ for the two wavelengths $\lambda=\SI{890}{nm}$ and $\lambda=\SI{970}{nm}$ at the 2nd antinode. Again the first red point corresponds to including all the main channels, and then background channels are added. For $\lambda=\SI{890}{nm}$, the initial propagating background radiation provides a significant increase to the efficiency, while the evanescent modes provide no change. On the other hand, for $\lambda=\SI{970}{nm}$, the initial propagating background radiation provides a significant decrease to the efficiency, while the evanescent modes provide a positive increase to the efficiency. This also showcases the complex interplay between the main channels and the background channels. 

\section{Perspective} \label{Sec:Discussion}

We have shown that contributions from multiple scattering channels influence both the Purcell factor and especially the collection efficiency. Unlike traditional Fabry-Pérot cavities, where scattering of the light is viewed simply as a loss mechanism, this scattering is in fact beneficial for the performance of the nanopost SPS. Importantly, this scattering mechanism decouples the efficiency from the Purcell factor directly challenging a well-known design paradigm that maximum collection efficiency is obtained on resonance. The identification of this mechanism opens a door to unconventional SPS design approaches, especially in the non-resonant regime where the scattering coefficients are no longer analysed and optimized with respect to the fundamental HE$_{11}$ mode alone, and where definitions of fundamental performance parameters such as the spontaneous emission $\beta$ factor need to be revisited.

For the nanopost itself, this invites to a new optimization of the collection efficiency with respect to all geometrical parameters. The maximum collection efficiency will be obtained for a reduced Purcell factor due to the new trade-off between efficiency and Purcell enhancement. 
Potential future work on the nanopost design could also be to explore the properties of cavity modes with different orders than 3. Additionally, structuring the bottom mirror could also lead to increased performance. Adding rings around the nanopost could positively alter the scattering mechanism while also bridging the gap to the closely related bullseye design \cite{Yao2018, Liu2019, Wang2019b}, which also features broadband collection efficiency independently of the Purcell factor. Despite flourishing literature on the bullseye design, the physical mechanisms underlying the performance is still unclear. The analysis of the bullseye will be more challenging as the inner mesa/nanowire features larger diameters resulting in additional guided modes. The rings around the inner mesa will also heavily influence the radiation modes and their mode profiles. This will, in turn, lead to changes in the emission rates and the reflection matrices and, thus, the scattering channels. In typical bullseye structures\cite{Yao2018, Liu2019, Wang2019b}, which are numerically optimized, the silica layer is hundred of nanometers thick. Here we anticipate that such large thicknesses will also result in increased mode coupling as light diverges when propagating in the silica.            
%What are the implications of our results? Outlook towards the bullseye design?

%Even better efficiency can be obtained by performing for non-resonant 

\section{Conclusions}

We have shown that the traditional Fabry-Pérot single-mode model, which typically provides an excellent description \cite{Friedler2009,Gregersen2016,Wang2020_PRB_Biying} of the physics for cavity-based single-photon sources, significantly underestimates the achievable performance of the nanopost structure. 
Using a modal expansion method, we have performed a detailed analysis of the emission channels. We have shown that in particular the collection efficiency benefits significantly from a contribution from light scattered to radiation modes, which often is simply considered a loss mechanism. 
This scattering into radiation modes not only allows for improved collection efficiency but also decouples the collection efficiency from the Purcell factor, such that optimum performance is obtained off-resonance. Our parameter scan of the nanopost structure reveals an achievable Purcell factor $F_{\rm p}$ of 7.9 or a collection efficiency $\varepsilon$ of $0.69$ obtained for two very different parameter sets. Our work invites further exploration of unconventional SPS design mechanisms, especially in the non-resonant regime.

\section*{Conflicts of interest}
There are no conflicts to declare.

\section*{Acknowledgements}
We thank Battulga Munkhbat for his assistance in creating the sketch of the nanopost. This work is funded by the European Research Council (ERC-CoG “UNITY,” Grant No. 865230), the French National Research Agency (Grant No. ANR-19-CE47-0009-02), the European
Union’s Horizon 2020 Research and Innovation Programme under the Marie Skłodowska-Curie Grant (Agreement No. 861097), and by the Independent Research Fund Denmark (Grant No. DFF-9041-00046B).

%%%END OF MAIN TEXT%%%


\balance

\renewcommand\refname{References}

%%%REFERENCES%%%
\bibliography{rsc} %You need to replace "rsc" on this line with the name of your .bib file
\bibliographystyle{rsc} %the RSC's .bst file

\newpage
\section{Supplementary}

\subsection{Height adjustment procedure} \label{heightpro}

In principle, the dipole position $h_{\rm t}$ and $h_{\rm b}$ would also be free parameters if we wanted to perform a complete optimization of the nanocavity. However, this would be a very demanding task. Instead, we have chosen to study cavities with 3 antinodes as our test simulations have shown that the 2nd antinode in a 3 antinode cavity provides a good performance. Thus $h_{\rm t}$ and $h_{\rm b}$ will be fixed to ensure a cavity with 3 antinodes and a resonance wavelength of $\lambda_{\mathrm{r}}=930\mathrm{nm}$ at the 2nd antinode. A procedure is required such that $h_{\rm t}$ and $h_{\rm b}$ satisfies these conditions. The first step is to use the SMM (phase conditions of the fundamental mode) to determine the initial total height $h$:  

%thus we will fix $h_{\rm t}$ and $h_{\rm b}$ using the following method: First, we decide that we wish to have a total of 3 antinodes in the cavity. The reason for this is, that our test simulations have shown that the second antinode in a 3 antinode cavity has good performance. The second step is to use a 1D model (phase conditions of the fundamental mode) to determine the initial total height $h$:

\begin{equation}
	h=(2\pi-\arg(r_{\rm bot,11}))/(2\beta_{1})+(2\pi-\arg(r_{\rm top,11}))/(2\beta_{1}).
\end{equation} This is under the conditions $\arg(r_{\rm bot,11})<0$ and $\arg(r_{\rm top,11})>0$. Then we place the dipole in the second antinode from the bottom:

\begin{equation}
	h_{\rm b}=(2\pi-\arg(r_{\rm bot,11}))/(2\beta_{1})
\end{equation}
and
\begin{equation}
	h_{\rm t}=(2\pi-\arg(r_{\rm top,11}))/(2\beta_{1}).
\end{equation}
However, due to the background continuum and mode-coupling, this method does not ensure that the dipole is placed exactly at an antinode nor that the resonant wavelength of the cavity corresponds exactly to the design wavelength, $(\frac{\mathrm{d}F_{\rm p}}{\mathrm{d}\lambda})_{\lambda=\lambda_d}=0$. To solve this problem, the dipole is first adjusted to the exact position of the antinode by plotting $|E_r(z)|^2$ and locating the peak. Then the height is slightly adjusted $h=h\pm\delta h$, while the dipole is moved to the exact position of the antinode for each adjustment until $(\frac{\mathrm{d}F_{\rm p}}{\mathrm{d}\lambda})_{\lambda=\lambda_d}=0$ is satisfied. Finally, the position of the 1st antinode (from the bottom) can also be identified by plotting $|E_r(z)|^2$.

Finally, we will provide the total height, $h_{\rm total}$ of the structure along with the height deviation between the initial height obtained from the SMM and the final height, $h_{\rm diff}=h_{\rm total}-h_{\rm initial}$.

%\begin{figure}[H]
	%\advance\leftskip-4cm
%	\begin{subfigure}{0.25\textwidth}
%		\centering
		%\includegraphics[width= 1 \textwidth]{GaAsnanoDBR2.eps}
%		\includegraphics[width= 1 \textwidth]{htotal.eps}
%		\caption{}
%		\label{htotal}
%	\end{subfigure}
%	\begin{subfigure}{0.25\textwidth}
%		\centering
		%\includegraphics[width= 1 \textwidth]{GaAsnanoDBR2.eps}
%		\includegraphics[width= 1 \textwidth]{hdiff.eps}
%		\caption{}
%		\label{hdiff}
%	\end{subfigure}
%	\caption{(a) Total height of the structure. (b) The height difference between the initial height and the final height.}
%\end{figure}

\begin{figure}[H]
	%\advance\leftskip-4cm
	\begin{subfigure}{0.5\textwidth}
		\centering
		\includegraphics[width= 1 \textwidth]{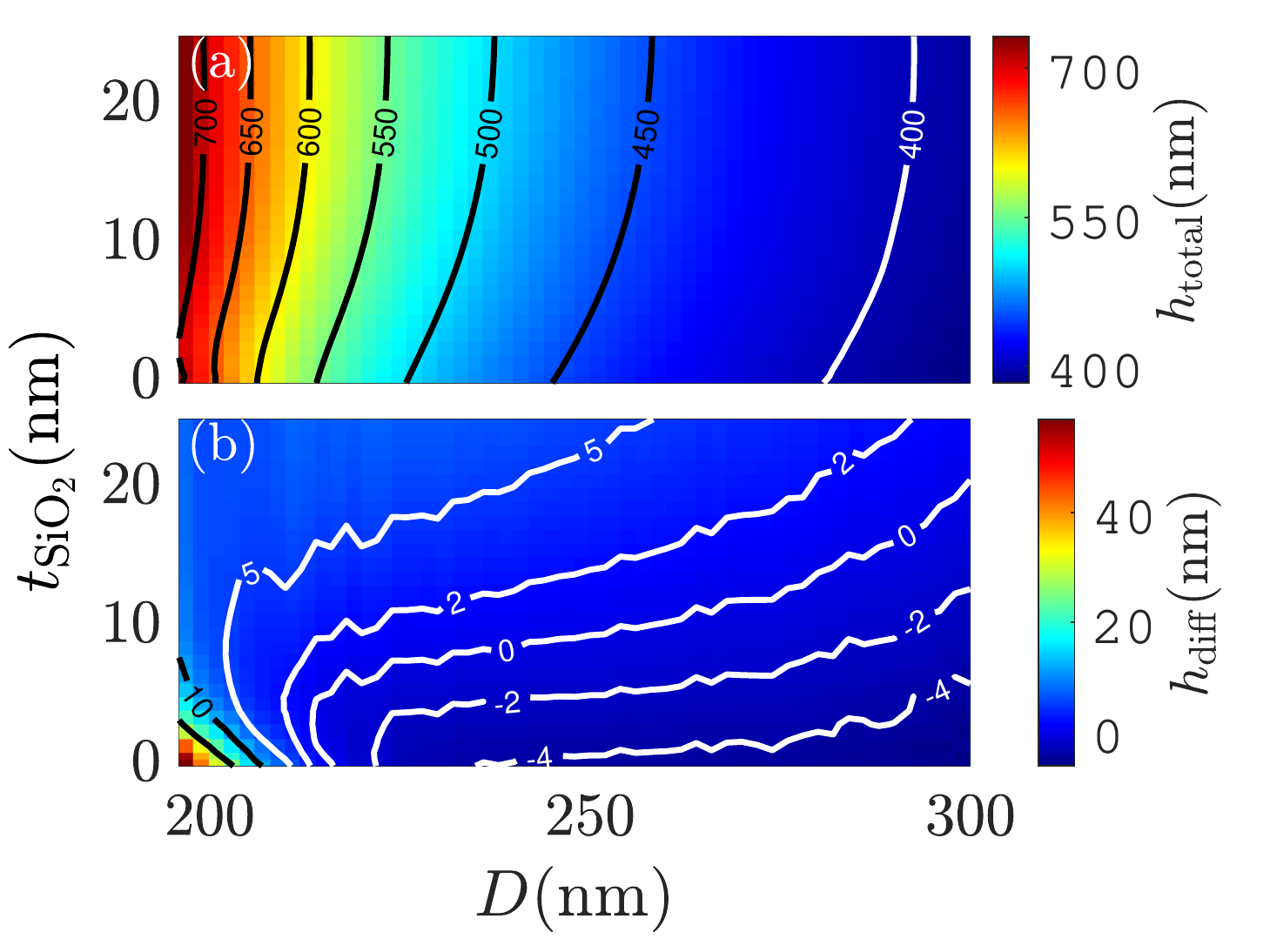}
		%\caption{}
	\end{subfigure}
	\caption{(a) Total height of the structure, $h_{\rm total}$, as a function of the diameter, $D$, and the silica layer thickness, $t_{\rm SiO_2}$. (b) The height difference between the initial height and the final height, $h_{\rm diff}$, as a function of the diameter, $D$, and the silica layer thickness, $t_{\rm SiO_2}$.}
	\label{height}
\end{figure}

In Fig.\ (\ref{height}a), the total height of the structure is shown as a function of the diameter and the silica layer thickness. The total height mostly depends on the diameter as the diameter determines the propagation constant $\beta_{1}$ and thus the primary influence of the phase. In Fig.\ (\ref{height}b), the height difference of the final total height compared to the SMM is shown. For most of the parameters, the difference is small in the range $\SI{-5}{nm}$ to $\SI{5}{nm}$. However, for the small diameters and thin silica layer thicknesses, there is a very large difference up to $\SI{50}{nm}$. This is the same parameter region where the modal reflection at the bottom interface is small. Thus the phase of the fundamental mode is less dominating compared to the contributions of the radiation and evanescent modes. Oscillations can also be observed in the height difference due to numerical noise. The exact resonance wavelength is sensitive to the height of the structure, but these oscillations are on the scale of less than $\SI{1}{nm}$, and the uncertainty in the resonance wavelength will be on a similar scale.   

\subsection{Influence of the numerical aperture} \label{sec:NA}

In all the simulations of the efficiency a numerical aperture of $\mathrm{NA}=0.75$ has been used. Here we will investigate the influence of varying the numerical aperture.

%\begin{figure}[H]
	%\advance\leftskip-4cm
%	\begin{subfigure}{0.25\textwidth}
%		\centering
		%\includegraphics[width= 1 \textwidth]{GaAsnanoDBR2.eps}
%		\includegraphics[width= 1 \textwidth]{eff2nd_NA04.eps}
%		\caption{}
%		\label{eff2nd_NA04}
%	\end{subfigure}
%	\begin{subfigure}{0.25\textwidth}
%		\centering
%		%\includegraphics[width= 1 \textwidth]{GaAsnanoDBR2.eps}
%		\includegraphics[width= 1 \textwidth]{eff2nd_NA1.eps}
%		\caption{}
%		\label{eff2nd_NA1}
%	\end{subfigure}
%	\begin{subfigure}{0.25\textwidth}
%		\centering
%		%\includegraphics[width= 1 \textwidth]{GaAsnanoDBR2.eps}
%		\includegraphics[width= 1 \textwidth]{eff2nd_curve_NA.eps}
%		\caption{}
%		\label{eff2nd_curve_NA}
%	\end{subfigure}
%	\begin{subfigure}{0.25\textwidth}
%		\centering
%		%\includegraphics[width= 1 \textwidth]{GaAsnanoDBR2.eps}
%		\includegraphics[width= 1 \textwidth]{eff_D238_t0_lambda_NA.eps}
%		\caption{}
%		\label{eff_D238_t0_lambda_NA}
%	\end{subfigure}
%	\caption{Efficiency at the 2nd antinode for a numerical aperture of $NA=0.4$ (a) and $NA=1.00$ (b). (c) Efficiency as a function of $NA$ for three different parameters at the 2nd antinode. (d) Efficiency as a function of wavelength for four different values of the $NA$.}
%	\label{eff_NA}
%\end{figure}

\begin{figure}[H]
	%\advance\leftskip-4cm
	\begin{subfigure}{0.5\textwidth}
		\centering
		\includegraphics[width= 1 \textwidth]{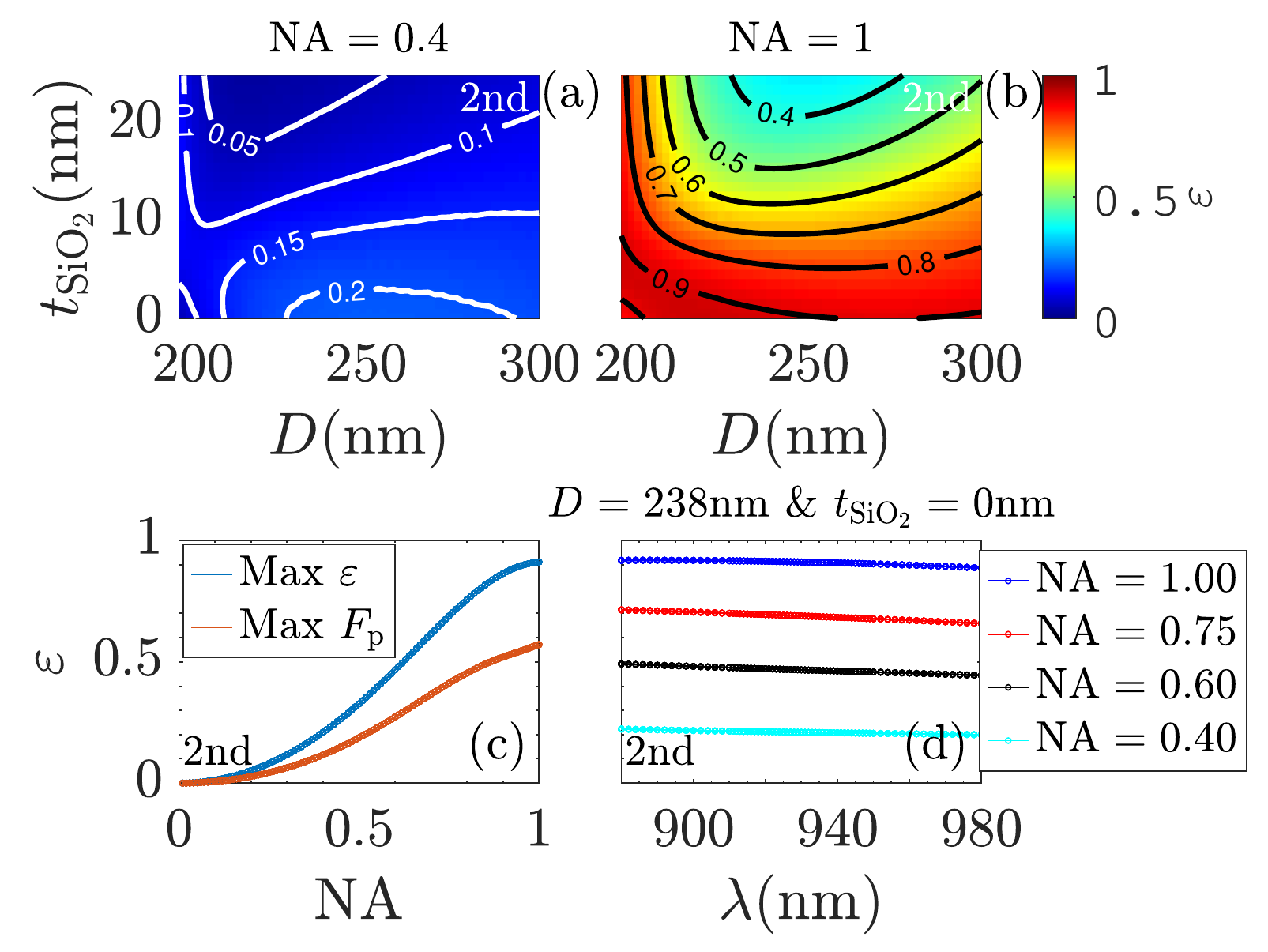}
	\end{subfigure}
	\caption{Efficiency, $\varepsilon$, at the 2nd antinode for a numerical aperture of $\mathrm{NA}=0.4$ (a) and $\mathrm{NA}=1.00$ (b). (c) Efficiency, $\varepsilon$, as a function of the numerical aperture, $\mathrm{NA}$, for three different parameters at the 2nd antinode. (d) Efficiency, $\varepsilon$, as a function of wavelength, $\lambda$, for four different values of the $\mathrm{NA}$.}
	\label{eff_NA}
\end{figure}

In Fig.(\ref{eff_NA}a) and Fig.(\ref{eff_NA}b) the efficiency is shown as a function of the diameter and the silica layer thickness similar to Fig.\ (\ref{eff_result}a), but for a numerical aperture $\mathrm{NA}=0.4$ and $\mathrm{NA}=1$. Lowering the numerical aperture to $\mathrm{NA}=0.4$ drastically reduces the efficiency and the maximum is barely above $\varepsilon=0.2$. This shows that a large numerical aperture is crucial for the good performance of the nanopost. By increasing the numerical aperture from $\mathrm{NA}=0.75$ ($\theta\approx49^{\circ}$) to $\mathrm{NA}=1.00$ ($\theta=90^{\circ}$) there is roughly a $20\%$ increase in the efficiency, so there is still some light lost at angles above $\theta\approx49^{\circ}$. Furthermore, for a numerical aperture of $\mathrm{NA}=1$, the efficiency directly represents the losses to the bottom mirror. For diameters above $D=\SI{210}{nm}$, an increased silica layer thickness increases the losses to the bottom mirror even though the Purcell factor increases.
In Fig.\ (\ref{eff_NA}c) the efficiency for the structures with the largest efficiency and Purcell factor, at the 2nd antinode, are shown as a function of the numerical aperture. The steepest part of the curves is roughly in the interval $\mathrm{NA}=0.4$ to $\mathrm{NA}=0.75$, which is the reason for the huge difference in efficiency between $\mathrm{NA}=0.4$ and $\mathrm{NA}=0.75$. The curves also start to flatten out as the $\mathrm{NA}$ reaches $1$. 
In Fig.\ (\ref{eff_NA}d) the efficiency is plotted for four different values of the numerical aperture as a function of the wavelength. The numerical aperture does not influence the curvature of the efficiency as a function of the wavelength. This means that being on resonance does not focus the far-field compared to being off resonance.    

\subsection{Gaussian collection efficiency} \label{sec:Gauss}

\begin{figure}[tb]
	%\advance\leftskip-4cm
	\begin{subfigure}{0.5\textwidth}
		\centering
		\includegraphics[width= 1 \textwidth]{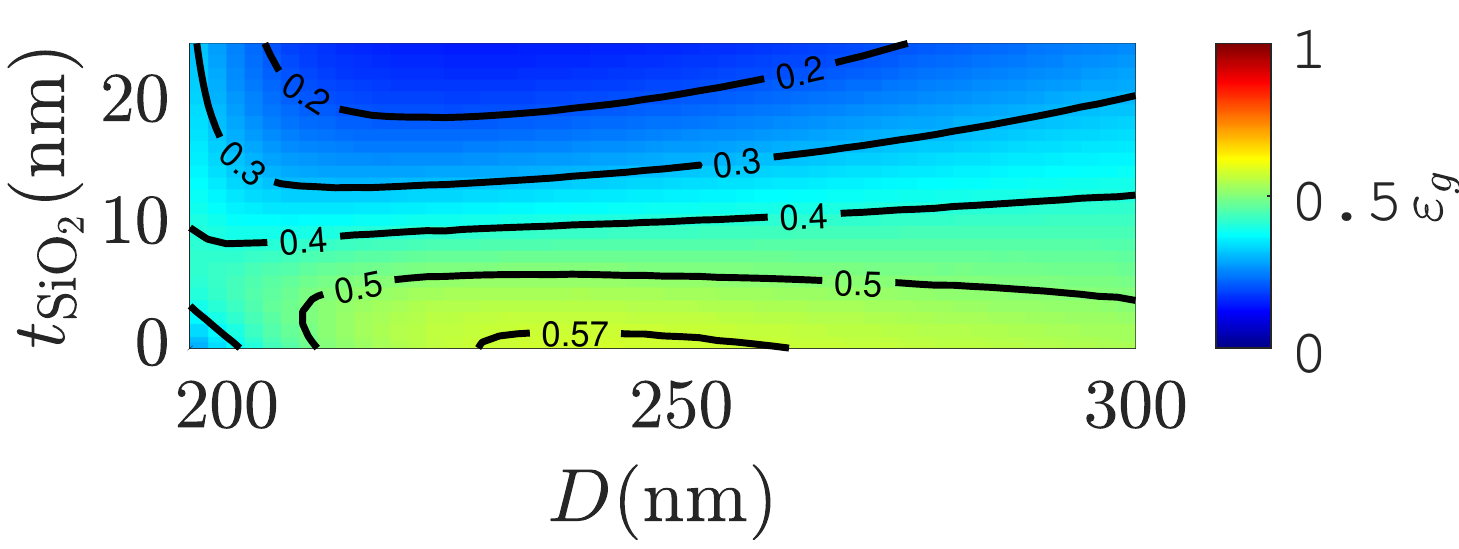}
	\end{subfigure}
	\caption{Gaussian collection efficiency, $\varepsilon_g$, as a function of the diameter, $D$, and the silica layer thickness, $t_{\rm SiO_2}$, for the 2nd antinode.}
	\label{eff2nd_gauss}
\end{figure}

So far the efficiency has been evaluated by calculating the total power collected in the lens with some numerical aperture. However, in many applications the light will couple to a fiber afterwards. Therefore we have also calculated the power overlap between the emitted far-field and the far-field of a Gaussian representative for the fundamental mode in many single-mode fibers \cite{Munsch2013}. The applied method is identical to the one presented in the appendix of \cite{Wang2020_PRB_Biying}, and the Gaussian collection efficiency is defined as $\varepsilon_g=P_{\rm collected, Gaussian}/P_{\rm T}$, where $P_{\rm collected, Gaussian}$ is defined as the overlap with a Gaussian profile.
In Fig.\ (\ref{eff2nd_gauss}) the Gaussian efficiency is shown for the 2nd antinode. Compared to the standard efficiency in Fig.\ (\ref{eff_result}a), the difference is approximately $0.1$ over the entire parameter space, showcasing the Gaussian shaped profile of the far-field.

\subsection{Efficiency analysis for the structure with maximum Purcell factor} \label{sec:eff max Fp}

We will now apply the efficiency analysis for the structure with the largest Purcell factor with the parameters $D=\SI{250}{nm}$ and $t_{\rm SiO_2}=\SI{13}{nm}$ and an efficiency of $\varepsilon=0.41$.

%\begin{figure}[H]
	%\advance\leftskip-4cm
%	\begin{subfigure}{0.25\textwidth}
%		\centering
		%\includegraphics[width= 1 \textwidth]{GaAsnanoDBR2.eps}
%		\includegraphics[width= 1 \textwidth]{efficiency_initial_D246_t12_NA75.eps}
%		\caption{}
%		\label{efficiency_initial_D246_t12_NA75}
%	\end{subfigure}
%	\begin{subfigure}{0.25\textwidth}
%		\centering
		%\includegraphics[width= 1 \textwidth]{GaAsnanoDBR2.eps}
%		\includegraphics[width= 1 \textwidth]{efficiency_final_D246_t12_NA75.eps}
%		\caption{}
%		\label{efficiency_final_D246_t12_NA75}
%	\end{subfigure}
%	\caption{(a) $\varepsilon$ ($NA=75$) as a function of the initial coefficients expressed with the propagation constant $\beta$. (b) $\varepsilon$ ($NA=75$) as a function of the final coefficients expressed with the propagation constant $\beta$.}
%	\label{efficiency_D246_t12_NA75}
%\end{figure}

\begin{figure}[H]
	%\advance\leftskip-4cm
	\begin{subfigure}{0.5\textwidth}
		\centering
		\includegraphics[width= 1 \textwidth]{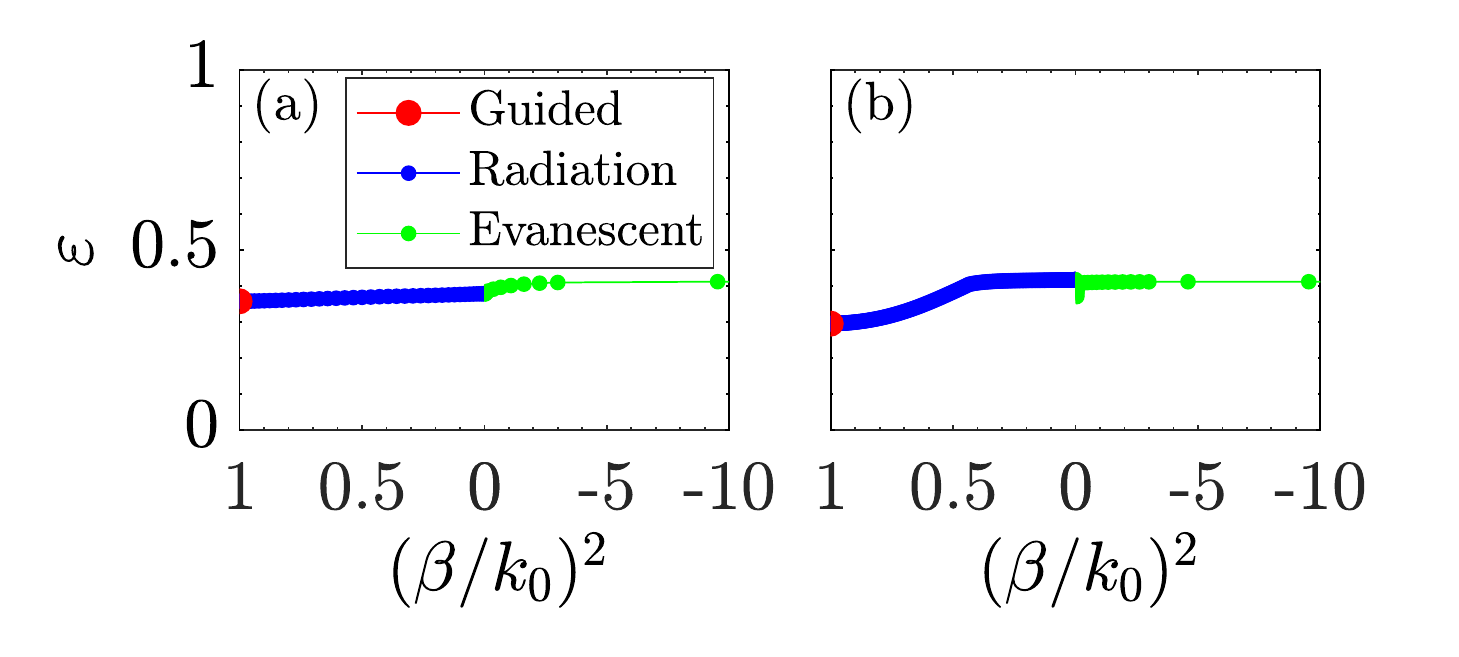}
	\end{subfigure}
	\caption{(a) Efficiency, $\varepsilon$ ($\mathrm{NA}=0.75$), as a function of the initial coefficients expressed with the propagation constant $(\beta/k_0)^2$. (b) Efficiency, $\varepsilon$ ($NA=75$), as a function of the final coefficients expressed with the propagation constant $(\beta/k_0)^2$.}
	\label{eff_D250_t13_model_beta2}
\end{figure}

In Fig.\ (\ref{eff_D250_t13_model_beta2}) the efficiency is shown as a function of the initial coefficients (Fig.\ (\ref{eff_D250_t13_model_beta2}a)) and the final coefficients (Fig.\ (\ref{eff_D250_t13_model_beta2}b)), expressed with the propagation constant $(\beta/k_0)^2$, just as for the structure with maximum efficiency. Again the curve in Fig.\ (\ref{eff_D250_t13_model_beta2}a) is flat and the channels of the fundamental mode dominates the efficiency. The efficiency increase by adding the final coefficients, i.e. $c_2$ and $c_3$, seen in Fig.\ (\ref{eff_D250_t13_model_beta2}b), is still significant, but much smaller compared to the structure with maximum efficiency. Here the increase is from approximately $\varepsilon=0.3$ to $\varepsilon=0.41$. The curve in Fig.\ (\ref{eff_D250_t13_model_beta2}b) also flattens out completely due to the numerical aperture.

\begin{figure}[H]
	%\advance\leftskip-4cm
	\begin{subfigure}{0.5\textwidth}
		\centering
		\includegraphics[width= 1 \textwidth]{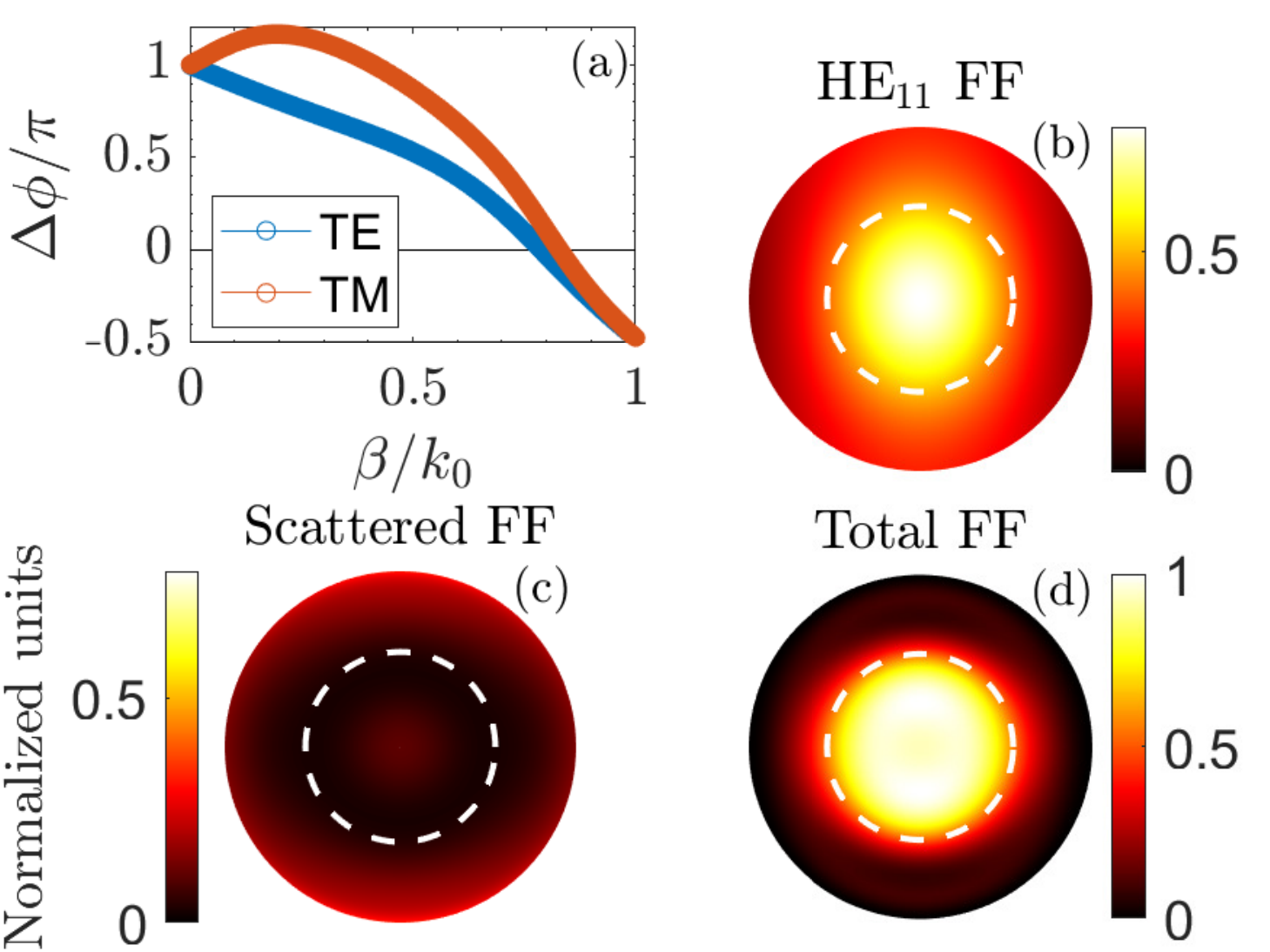}
	\end{subfigure}
	\caption{(a) The phase difference between the direct transmission of the fundamental mode and the background continuum for TE and TM modes as a function of the propagation constant. (b) The far-field of the fundamental mode. (c) The far-field of the background continuum. (d) The total far-field. The white dotted line indicates $\mathrm{NA}=0.75$. Be aware of the different color scales that have been used for the far-fields.}
	\label{farfield_D250_t13}
\end{figure}

In Fig.\ (\ref{farfield_D250_t13}a) the  phase difference in the air layer between the direct transmission of the fundamental mode and the entire background is shown as a function of the propagation constant for TE and TM modes. We observe similar features as before, i.e. constructive (destructive) interference for light propagating vertically (horizontally). Though at $\beta/k_0=1$, the phase difference is larger compared to the previous structure. In Fig.\ (\ref{farfield_D250_t13}b), Fig.\ (\ref{farfield_D250_t13}c) and Fig.\ (\ref{farfield_D250_t13}d) the far-fields of the direction transmission of the fundamental mode, the background radiation and the total field is shown. Compared to the structure with maximum efficiency, the far-field of the background radiation is significantly different. Here, the far-field is mainly focused towards horizontal angles and the intensity is much smaller compared to the far-field of $\mathrm{HE}_{11}$. As such the constructive contribution at smaller angles is not as significant and less of the radiation will be captured by the lens, due to the numerical aperture. This explains why the efficiency increase in Fig.\ (\ref{eff_D250_t13_model_beta2}b) is much smaller compared the structure with the maximum efficiency. However, there is still destructive interference for the light that propagates horizontally. As such the interference between the direction emission and the radiation focuses the far-field, but not to the same degree as for the structure with the maximum efficiency.

\subsection{Purcell factor analysis for the structure with maximum collection efficiency} \label{sec:Fp max eff}

We will now apply the model for the Purcell factor for the structure with the largest efficiency, $D=\SI{238}{nm}$ and $t_{\rm SiO_2}=\SI{0}{nm}$ and $F_\mathrm{p}=4.8$.

\begin{figure}[h]
	%\advance\leftskip-4cm
	\begin{subfigure}{0.5\textwidth}
		\centering
		\includegraphics[width= 1 \textwidth]{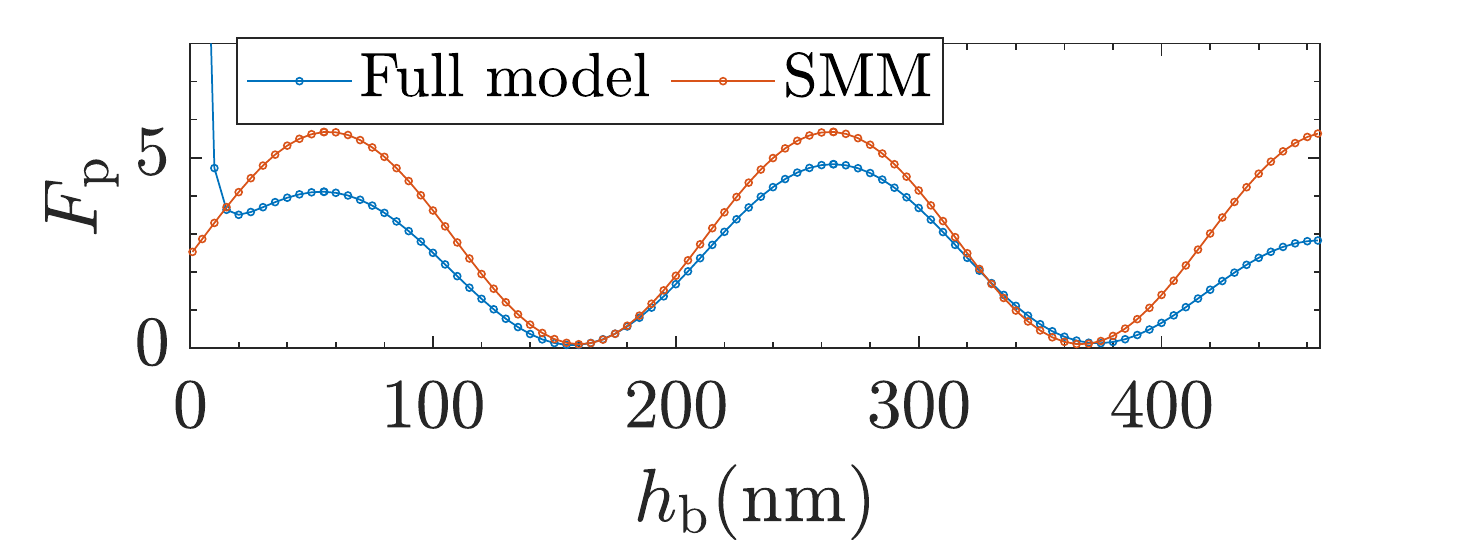}
	\end{subfigure}
	\caption{Purcell factor, $F_{\rm p}$, computed using the full model and the SMM as a function of the dipole position from the bottom interface, $h_{\rm b}$. $D=\SI{238}{nm}$ and $t_{\rm SiO_2}=\SI{0}{nm}$.}
	\label{Fp_D238_t0_hb}
\end{figure}

In Fig.\ (\ref{Fp_D238_t0_hb}) the Purcell factor is shown as a function of the dipole position throughout the cavity, both the full model (n. 7) and the SMM (n. 1) are used. Here the 3 antinodes can be observed and the SMM predicts a larger Purcell factor for all 3 antinodes compared to the full model. The positions of the antinodes are almost identical between the full model and the SMM. The Purcell factor increases drastically when the dipole is placed close to the metal mirror due to non-radiative decay processes\cite{novotny2012principles}.

\begin{figure}[h]
	%\advance\leftskip-4cm
	\begin{subfigure}{0.5\textwidth}
		\centering
		\includegraphics[width= 1 \textwidth]{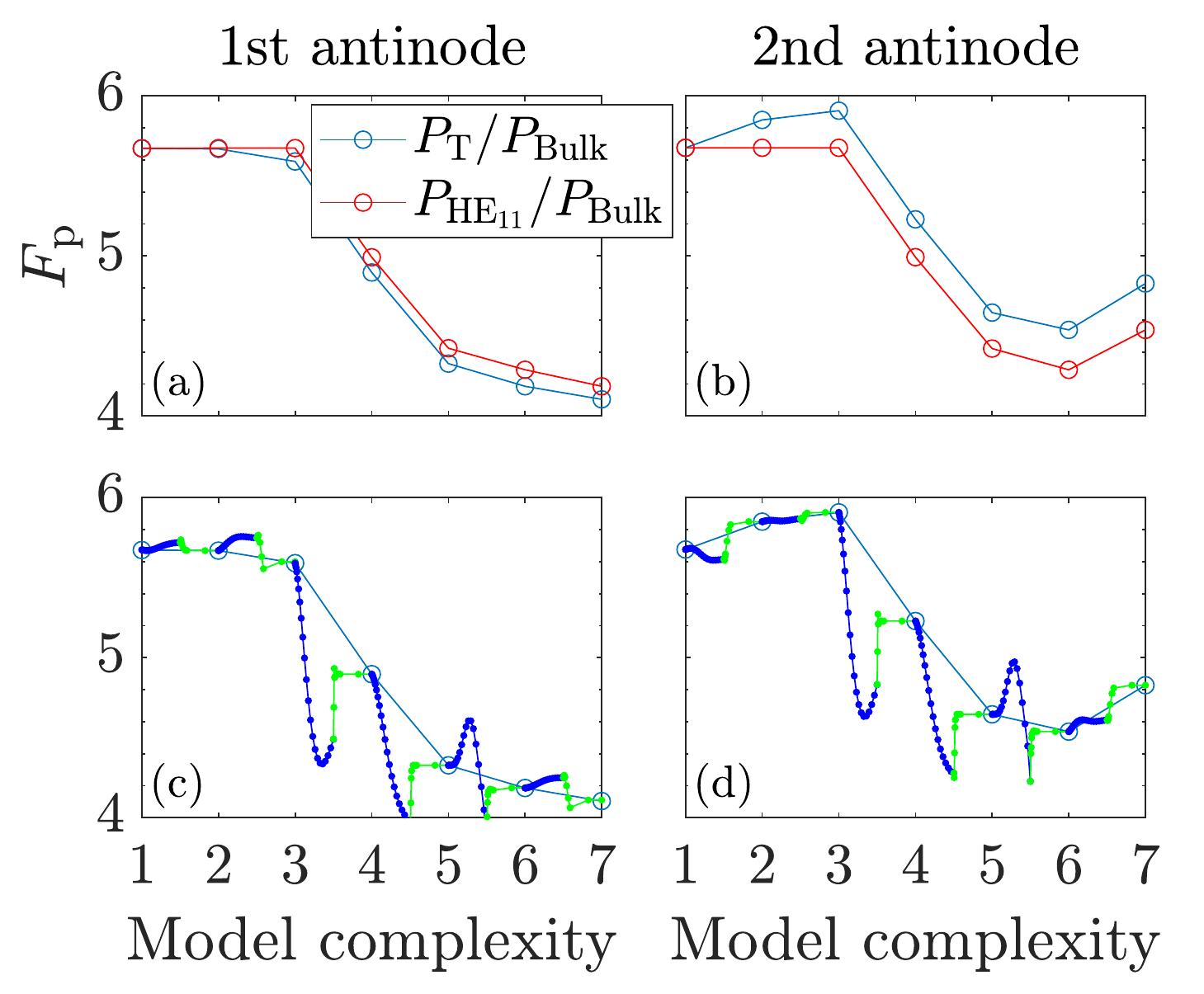}
	\end{subfigure}
	\caption{Purcell factor ($P_{\rm T}/P_{\rm Bulk}$) and fundamental mode enhancement ($P_{\mathrm{HE}_{11}}/P_{\rm Bulk}$) for the 1st (a) and 2nd (b) antinode as a function of the model number. In (c) and (d) the background continuum is continuously included between each model number.}
	\label{Fp_models_d238_t0}
\end{figure}

In Fig.\ (\ref{Fp_models_d238_t0}a) and Fig.\ (\ref{Fp_models_d238_t0}b) the Purcell factor and the power enhancement of the fundamental mode is shown as a function of the model complexityfor the 1st and 2nd antinode. Compared to the structure with $D=\SI{250}{nm}$ and $t_{\rm SiO_2}=\SI{13}{nm}$, there are a few differences. The SMM predicts a smaller Purcell factor, which is simply caused by the lower modal reflection at the bottom. There is a large negative contribution when including the back-scattering at the bottom interface (n 5.) at both antinodes. This is caused by the change of the silica layer thickness and as seen in Fig.\ (\ref{Fp_models_d238_t0}c) and Fig.\ (\ref{Fp_models_d238_t0}d). The propagating radiation modes are responsible for this decrease. Furthermore, by including the scattering of the background to itself (n. 6), there is now a small decrease for both antinodes. These are the differences between the two structures. The differences between the 1st and 2nd antinode are exactly the same for the two structures, where the scattering into evanescent modes at the top interface (n. 2) and the initial evanescent modes provide a positive contribution at the 2nd antinode. 

\subsection{Asymmetric wavelength dependence for the two antinodes} \label{sec:asym}
%(Should this be included??)\newline
To further study the resonance shift between the 1st and 2nd antinodes, we choose a nanopost design of $D=\SI{202}{nm}$ and $t_{\rm SiO_2}=\SI{5}{nm}$, where the shift is more pronounced. 

%During the parameter optimization/investigation a peculiar effect was encountered, namely an anomalous behaviour of the 1st and 2nd antinode in the nanowire optical nanocavity with small diameters. The resonance wavelength shifts depending on whether the dipole is placed in the 1st or 2nd antinode. To showcase this behaviour, we will present the Purcell factor for the two antinodes for a structure with $D=\SI{202}{nm}$ and $t_{\rm SiO_2}=\SI{5}{nm}$.

\begin{figure}[h]
	%\advance\leftskip-4cm
	\begin{subfigure}{0.5\textwidth}
		\centering
		\includegraphics[width= 1 \textwidth]{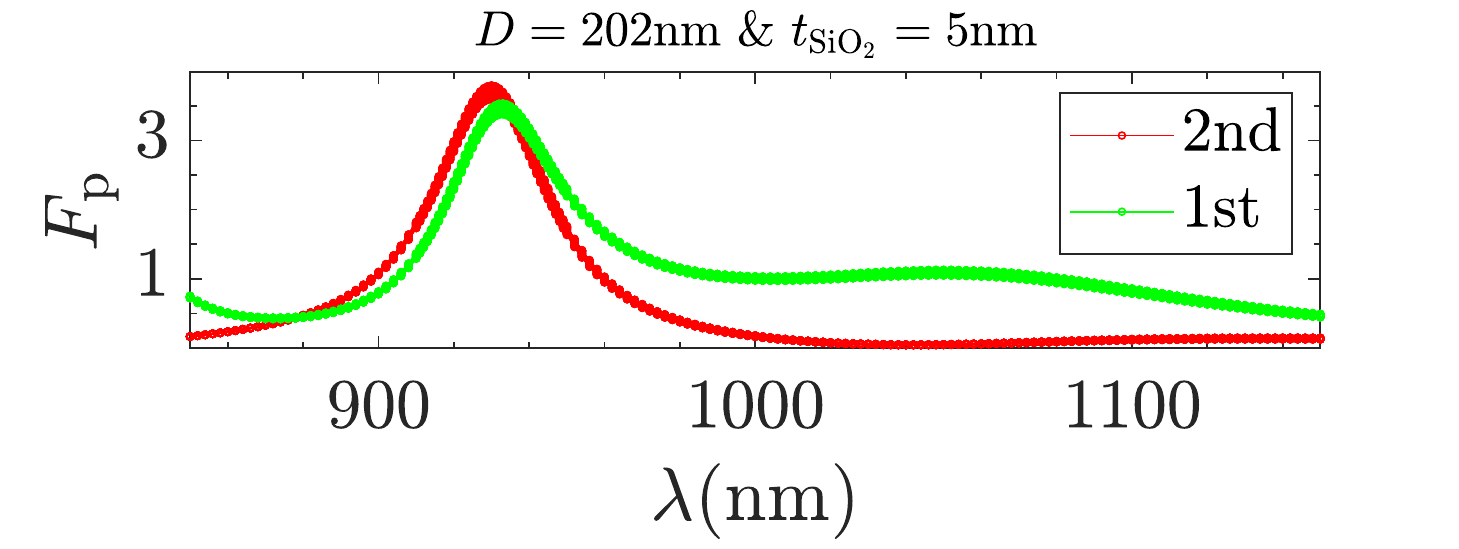}
	\end{subfigure}
	\caption{Purcell factor, $F_{\rm p}$, as a function of wavelength, $\lambda$, for the two antinodes. The parameters are $D=\SI{202}{nm}$ and $t_{\rm SiO_2}=\SI{5}{nm}$.}
	\label{Fp_D202_t5_lambda}
\end{figure}

In Fig.\ (\ref{Fp_D202_t5_lambda}) the Purcell factor is shown for the two antinodes as a function of the wavelength. The peak positions of the Purcell factors (resonance wavelength) are $\lambda_{2nd,r}=\SI{930}{nm}$ and $\lambda_{1st,r}=\SI{933}{nm}$. By observing the curve for the 1st antinode, this shift is caused by another broad resonance at approximately $\lambda=\SI{1050}{nm}$. To gain further insight into the resonances of the structure and verify our results, we have performed a quasi-normal mode (QNM) simulation\cite{RosenkrantzdeLasson:15} of the nanopost. In this simulation 15 QNMs are found and the complex eigenfrequencies, $\Tilde{\omega}_{\mu}=\omega_{\mu}-\mathrm{i}\gamma_{\mu}$, of the 3 important QNMs are $\Tilde{\omega}_{\rm QNM_1}=\num{2.0237e15}-\mathrm{i}\num{4.4904e13} \mathrm{Hz}$, $\Tilde{\omega}_{\rm QNM_2}=\num{1.7595e15}-\mathrm{i}\num{1.4654e14} \mathrm{Hz}$ and $\Tilde{\omega}_{\rm QNM_3}=\num{2.2809e15}-\mathrm{i}\num{2.0703e13} \mathrm{Hz}$. The corresponding real parts of the complex wavelength are $\lambda_{\rm QNM_1}=\SI{930.3}{nm}$, $\lambda_{\rm QNM_2}=\SI{1063.2}{nm}$ and $\lambda_{\rm QNM_3}=\SI{825.8}{nm}$. The Q factors of the QNMs can also be calculated using $Q_{\mu}=\omega_{\mu}/(2\gamma_{\mu})$ \cite{RosenkrantzdeLasson:15}, and we obtain $Q_{\rm QNM_1}=22.5$, $Q_{\rm QNM_2}=6.0$ and $Q_{\rm QNM_3}=55.1$.

%For this resonance at $\lambda=\SI{1050}{nm}$ the position of the 1st antinode is closer to an antinode, whereas the position of the 2nd antinode is closer to a node. There also seems to be a resonance for smaller wavelengths, but this is further away. To verify the results, a QNM simulation has been used to reproduce the same results. In this simulation 15 QNMs are found and the eigenfrequencies of the 3 important QNMs are $\Tilde{\omega}_{QNM_1}=\num{2.0237e15}-i\num{4.4904e13} \mathrm{Hz}$, $\Tilde{\omega}_{QNM_2}=\num{1.7595e15}-i\num{1.4654e14} \mathrm{Hz}$ and $\Tilde{\omega}_{QNM_3}=\num{2.2809e15}-i\num{2.0703e13} \mathrm{Hz}$. The corresponding real parts of the wavelength is $\lambda_{QNM_1}=\SI{930.3}{nm}$, $\lambda_{QNM_2}=\SI{1063.2}{nm}$ and $\lambda_{QNM_3}=\SI{825.8}{nm}$.

%$\lambda_{QNM_1}=\SI{931}{nm}$, $\lambda_{QNM_2}=\SI{1071}{nm}$ and $\lambda_{QNM_3}=\SI{826}{nm}$.

\begin{figure}[h]
	%\advance\leftskip-4cm
	\begin{subfigure}{0.5\textwidth}
		\centering
		\includegraphics[width= 1 \textwidth]{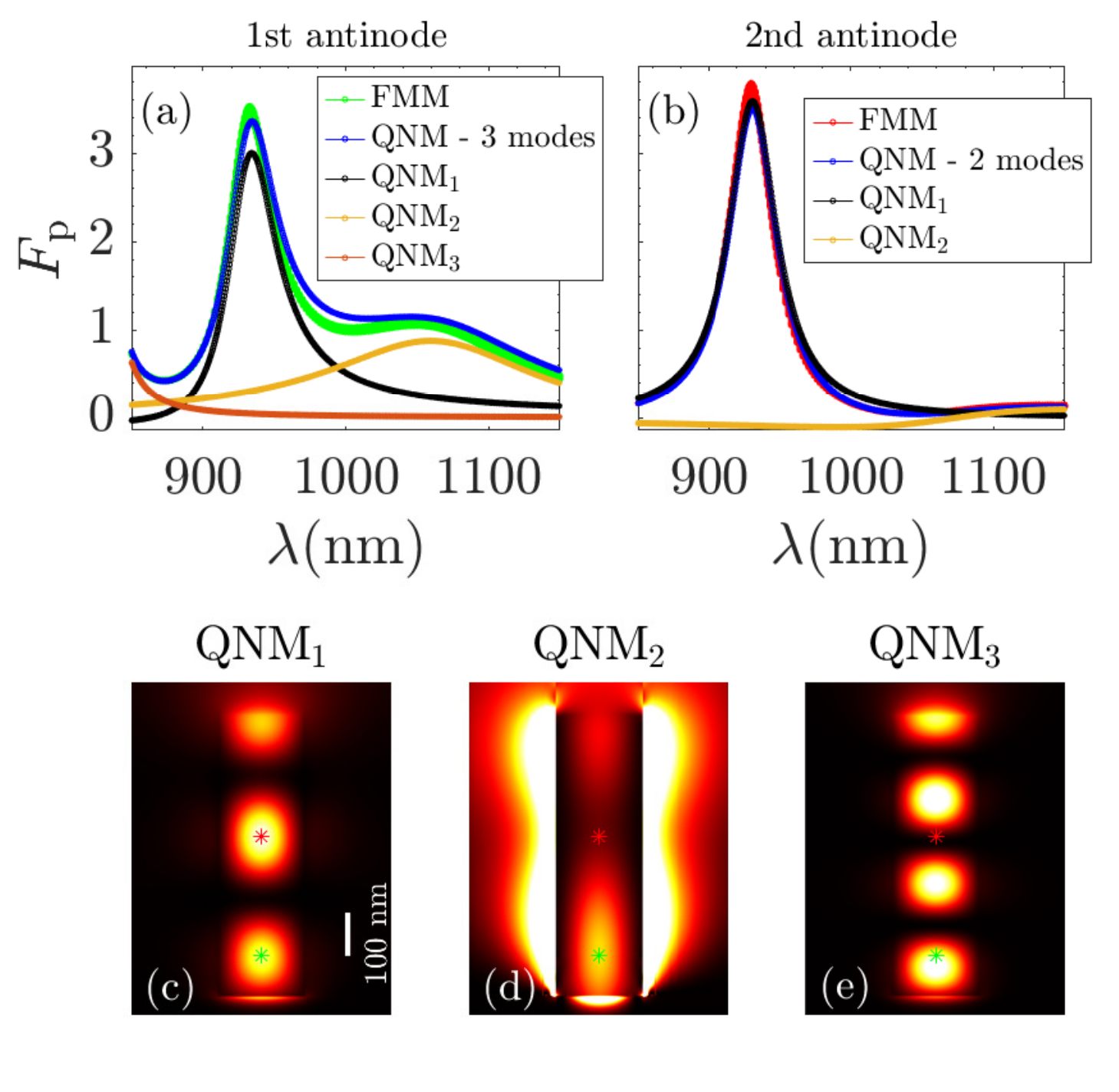}
	\end{subfigure}
	\caption{Comparison of the Purcell factor between the FMM and the QNM simulation for the 1st antinode (a) and the 2nd antinode (b). In-plane electrical field profiles of the 3 QNMs at their resonance wavelengths are shown in (c), (d) and (e). The green star corresponds to the position of the 1st antinode, and the red star corresponds to the position of the 2nd antinode. The white scale bar in (a) corresponds to \SI{100}{nm}. The intensity is scaled in each field plot and should not be used for comparison.}
	\label{QNM}
\end{figure}

In  Fig.\ (\ref{QNM}a,\ref{QNM}b), the comparison of the Purcell factor between the FMM and the QNM simulation is shown for the two antinodes. Overall, the quantitative agreement between the FMM and QNM simulations is good with some small deviations. In Fig.\ (\ref{QNM}a), the individual contributions of 3 QNMs are plotted along with their sum and the result of the FMM for the 1st antinode. These 3 QNMs provide a good description of the overall Purcell factor and they directly correspond to the peaks in the spectrum. QNM$_1$ and QNM$_2$ also overlap in the spectrum due to the low Q factor of QNM$_2$, which slightly shifts the peak position of the total Purcell factor. In Fig.\ (\ref{QNM}b), the individual contributions of 2 QNMs are plotted along with their sum and the result of the FMM for the 2nd antinode. Here $\rm QNM_1$ is almost sufficient to describe the entire spectrum, and we do not observe any other peaks than the one at $\lambda=\SI{930}{nm}$, besides a small bump at longer wavelengths. Now, consider the in-plane electrical field profiles of the 3 QNMs shown in Fig.\ (\ref{QNM}c,\ref{QNM}d,\ref{QNM}e). $\rm QNM_1$ has 3 antinodes, $\rm QNM_2$ has 2 antinodes and $\rm QNM_3$ has 4 antinodes. The green star corresponds to the position of the 1st antinode, where the QD is placed, and this position is very close to an antinode for $\rm QNM_2$ and $\rm QNM_3$. Therefore the contributions of these QNMs appear in the spectrum. However, the position of the 2nd antinode (red star) is much closer to a node for $\rm QNM_2$ and $\rm QNM_3$, and therefore they do not influence the spectrum.

%QNM$_1$ and QNM$_2$ also overlap in the spectrum which slightly shifts the resonance of the total Purcell factor. In Fig.\ (\ref{2nd_QNM}) the same simulation is shown for the 2nd antinode. Here QNM$_1$ is sufficient to describe most of the Purcell factor. Only at longer wavelengths QNM$_2$ should be included to describe the small bump. This means that QNM$_1$ has 3 antinodes corresponding to the positions found using the o-FMM. However, QNM$_2$ does not have 3 antinodes inside the cavity and the position of these antinodes are different. Then the 1st antinode is closer to one of these antinodes of QNM$_2$ than the 2nd antinode, which explains the difference in the spectrum.      

\end{document}